\documentstyle[12pt,bkqmb,epsf]{book}

\newfont{\bfa}{cmbx10 scaled\magstep1}
\def\leaderfill{\leaders\hbox to 1em{\hss.\hss}\hfill}

\def\third { {\textstyle {1\over 3}} }
\def\fourth{ {\textstyle {1\over 4}} }

\def\twelfth { {\textstyle {1\over 12}} }
\def\half{ {\textstyle {1\over 2}} }
\def\nsites {N_{\rm sites}}
\def\opvecm {\widehat{\vphantom{m^i}\smash{\overrightarrow m}}}
\def\vecm {\vphantom{m}\smash{\overrightarrow m}}
\def\xii {\vec\xi\,\vphantom{\xi}^i}
\def\cbar {\overline c}

\def\pmb#1  {\setbox0=\hbox{#1}%
    \kern-.025em\copy0\kern-\wd0
    \kern.05em\copy0\kern-\wd0
    \kern-.025em\raise.0433em\box0 }
\def\beq{\begin{equation}}
\def\eeq{\end{equation}}
\def\beqa{\begin{eqnarray}}
\def\eeqa{\end{eqnarray}}

\title {A New Approach to Doped Mott Insulators}
\author{N.E.\ Bickers$^*$\ and\ D.J.\ Scalapino$^{\dag}$ \\
        $\qquad$ }

\address{$^*$Department of Physics\\
         University of Southern California\\
         Los Angeles, CA 90089 \\
         $\qquad$            \\
         $^{\dag}$Department of Physics\\
         University of California\\
         Santa Barbara, CA 93106}

\begin{document}

\maketitle

\begin{abstract}
We describe a new microscopic approach for analyzing interacting electron
systems with local moments or, in principle, any local order parameter. We
specialize attention to the doped Mott insulator phase of the 
Hubbard model, where standard weak--coupling perturbation methods fail. A 
rotationally invariant Stratonovich--Hubbard field is introduced to decouple 
the static spin components of the interaction at each site. Charge degrees of freedom are 
then treated by ``slave'' Hartree--Fock in the presence of a spatially varying 
random spin field. The effective action reduces to a classical Heisenberg model 
at half filling, insuring that the system has (i)~finite--range order at $T>0$
with an exponentially diverging correlation length and (ii)~a one--electron Mott--Hubbard gap 
in the presence of disorder. Away from half--filling properties are determined by
strongly non--Gaussian fluctuations in the amplitude and orientation of the 
local spin fields. The saddle point equations of the theory at zero temperature
reduce to inhomogeneous Hartree--Fock, so that disordering of domain walls at finite
temperature is in principle included.  We present preliminary small system results for the 
intermediate and large interaction regimes obtained by Monte Carlo simulation of
random spin field configurations.

\end{abstract}

\vfill\eject

\section{Introduction}

In this paper we describe a new approach to perturbation theory for systems with
local moments or, in principle, any local order parameter. We begin by summarizing 
the present status of weak--coupling perturbation theories for such systems, with
the Hubbard model as a primary example.

Over the course of the last decade weak--coupling perturbation theories have been
used extensively to investigate the physics of the Hubbard model and its
extensions. The simplest perturbative approach to the Hubbard model is just Hartree--Fock
theory for one--body correlation functions. The conserving approximation implied for
two--body correlations is then the random phase approximation, or RPA. In the large--$U$
limit at half--filling the static spin response is unstable within Hartree--Fock, and the 
system exhibits a broken--symmetry ground state with infinite--range antiferromagnetic order \cite{swz}.
Furthermore a gap of order $U$ appears in the one--body density of states for the ordered
phase. If the Hartree--Fock equations are extended to finite temperature, long--range 
antiferromagnetic order onsets at a temperature of order $U$ in
the large--$U$ limit. The zero--temperature behavior of the Hartree--Fock approximation
away from half--filling has also been investigated, and incommensurate states, as well as
vertical and diagonal domain walls, have been studied \cite{zaanen}. However, the
appropriate treatment for finite temperatures has remained unclear, because the energetic
competition of the various states is delicate.

In fact the Hartree--Fock approximation at finite temperatures provides only a crude
description of the physics in a two-- (or one--) dimensional system. In accordance with
the Mermin--Wagner Theorem \cite{mermin}, infinite--range magnetic order is forbidden for $T>0$ in
two dimensions. The magnetic correlation length $\xi(T)$ instead diverges exponentially 
in $1/T$ for $T\rightarrow 0$ at half--filling, and the quantitative behavior of $\xi$,
as well as the zero--temperature ordered moment, is renormalized by quantum fluctuations.
Nevertheless the half--filled model exhibits a gap (or {\sl pseudogap}, since we do not
mean to imply a rigorously zero density of states) at finite temperature in the 
spin--disordered phase.

With the intention of modeling strong spin fluctuations in a disordered system, the present
authors proposed the fluctuation exchange, or FLEX, approximation \cite{bs,bsw,bw} as a 
plausible starting point.
The self--consistent treatment of spin fluctuations in the one--body self--energy suppresses
the magnetic ordering transition to zero temperature. Nevertheless the new approximation is
itself flawed: It does not produce a sharp gap at half--filling and instead smears out the
density of states, leaving a finite weight at zero energy for all temperatures \cite{bsw,bw,moukouri}. 
Furthermore the spin response function diverges with decreasing temperature in a way which only roughly
mirrors the exact behavior of the model. The failure at half--filling calls into question
results obtained for large $U$ at small doping, and in the end the FLEX approximation
fails to describe the pseudogap regime.

The inadequacy of the FLEX approximation has motivated various attempts to improve the theory. 
Kampf and Schrieffer have calculated the electron spectral density near an antiferromagnetic transition to
lowest order in an effective spin--fluctuation interaction, and their results exhibit a pseudogap when the
spin correlation length is large \cite{kampf}. Vilk and Tremblay \cite{vilk} have introduced a factorization
approximation for the electron four--point function which is constrained to be exact when all 
space--time coordinates coincide. Others \cite{chubukov1} have argued that perturbative calculations based on
an unrenormalized electron propagator are justified by a cancellation between vertex corrections and
self--energies, leading to the appearance of a pseudogap in some parameter regimes.

In addition efforts have been made to extend the FLEX approximation within a scheme based
on infinite--order (or ``parquet'') vertex renormalization \cite{dominicis,parquet0,parquet1}. 
Quantitative studies of parquet renormalization have had the goal of establishing the validity or inadequacy of
perturbation theory resummations for systems with a gap, but without infinite--range order.
Why would one assume that such resummations have any chance of succeeding?
For at least one nontrivial model, the Anderson impurity in a metallic host, it has been
known since the mid--1980's that zero--temperature perturbation theory for thermodynamic
properties has an infinite radius of convergence in $U$ \cite{zlatic}. This fact, as well as some
suggestive results in the low--density limit, initially encouraged hope that an infinite--order parquet
summation for the doped Mott insulator might succeed if the technical difficulties impeding 
its implementation could be overcome.

In fact this hope is not justified. Full numerical solutions of the minimal parquet 
equations for the two--dimensional
Hubbard model in the large--$U$ limit \cite{parquet2} yield the following results: (i)~No gap appears in 
the converged solution at half filling. (ii)~The spectral density at
low energies is actually larger than that in a FLEX calculation. We interpret these results to mean that
conventional perturbation theory \cite{kondo} fails to describe the physics of a Mott--Hubbard gap or
pseudogap in a spin--disordered system. Furthermore the parquet results, which consistently incorporate
vertex renormalization to infinite order, 
bring into question whether the basic problem can be circumvented by the constrained factorization
approximation or a cancellation between parts of perturbative diagrams.

Rather we believe the failure of perturbation theory around a uniform (i.e., Fermi liquid)
starting point is due to a qualitative change in the saddle--point structure of the
action for the Hubbard model at temperatures of order $U$ in the large--$U$ limit. This
change is induced by the appearance of local moments (or, more generally speaking,
a local order parameter field) in the theory. Conventional perturbation theory becomes
unreliable below a temperature scale which depends on both the interaction $U$ and the
chemical potential $\mu$. Below this scale one needs a revised perturbation theory which
incorporates ``anomalous'' Feynman graphs, but does so without inducing infinite--range 
order. These seemingly
contradictory objectives may be accomplished by introducing a spatially varying static
field with the full symmetry of the associated classical order parameter. One way to
introduce such a field is via a Stratonovich-Hubbard (S--H) transformation \cite{strat,hubb} 
of the interaction.
In particular we employ below a spin--rotation--invariant S--H field to decouple the static
$(m=0)$ component of the Hubbard interaction. About thirty years ago 
Wang et al. \cite{wang} showed for the magnetic alloy problem
that the effective static free energy induced by a time--varying S--H field provides
a physical picture of the crossover from band--like to local--moment behavior. 
Evenson et al. \cite{evenson} analyzed the coupling between isolated moments which arises
within this approach and noted that the simplest scheme generates an Ising, rather than
Heisenberg, interaction. Below we show that when a vector S--H field is introduced, the
effective action reduces to a classical Heisenberg form at half--filling.
This means no infinite--range order at $T>0$, an exponentially
diverging correlation length $\xi(T)$, {\sl and} a  Mott--Hubbard gap in the spin--disordered
phase. Furthermore the approximation scheme should provide a systematic way to study doping of
the Mott insulator. In the doped case both orientation and amplitude fluctuations of the field
become important. In particular the effective action for the static field is strongly
non--Gaussian \cite{millis}. Since the action arises directly from integrating out the fermion degrees 
of freedom, its form reflects both the underlying spin and charge structure of 
the system. In this way domain wall stripe fluctuations can occur naturally if
they are present in the model.

We describe this new approximation scheme for the Hubbard model in Section 2, then derive expressions for
specific electron correlation functions in Section 3. In Section 4 we provide a
brief description of our Monte Carlo sampling scheme for integration over the fluctuating
local fields. In Section 5 we present some illustrative results
obtained on a work station for a $4\times 4$ system. We outline in Section 6 the
form of generalized approximations which extend the present approach in the same sense that
FLEX extends Hartree--Fock. Finally we close in Section 7 with some speculations on possible implications for 
high--temperature superconductivity and with an outline of the next steps in this work.

\section{Approximation scheme}

In this section we show how to decouple the the static spin interaction in the repulsive
Hubbard model by introducing a spin--rotation--invariant S--H field.
This field may be interpreted as a classical Heisenberg order parameter which generalizes
the saddle--point in conventional temperature--dependent Hartree--Fock calculations.

It is conventional to decouple the interaction term by introducing 
independent fields at each point in imaginary time. When this is done, some freedom remains
in choosing the form of the interaction, due to the equal--time fermion anticommutation
relations. One may write equivalently
\beqa
U\widehat n_{i\uparrow}\widehat n_{i\downarrow}\ &=&\ \fourth U(\widehat n_i^2
  - \widehat m_{iz}^2) \label{intone1}\\ 
  &=&\ \half U(\widehat n_i - \widehat m_{iz}^2) \label{intone2}\\
  &=&\ \half U\widehat n_i -\fourth U(\widehat m_{ix}^2 + \widehat m_{iy}^2)\ , \label{intone3} 
\eeqa
where
\beqa
\widehat n_i\ &=&\ \widehat n_{i\uparrow} + \widehat n_{i\downarrow} \nonumber \\
\opvecm_i\ &=&\ \sum_{\alpha\alpha'}\,
  c^{\dag}_{i\alpha}\vec\sigma_{\alpha\alpha'}
  c^{\vphantom{\dag}}_{i\alpha'}\ . 
\label{opsone}
\eeqa
There is in fact considerably more freedom allowed, since the interaction is rotationally invariant: The
z--component of the moment operator in Eqns.~\ref{intone1} and \ref{intone2} may be
replaced by the x-- or y--component, and likewise different pairwise combinations of moment
operators may be substituted in Eqn.~\ref{intone3}. One is free to consider in addition arbitrary
linear combinations of these expressions, weighted properly to reproduce the original interaction.

This was realized by several authors around 1970 during early studies of the
single--impurity Anderson model and the Hubbard model. In particular Wang et al. employed 
Eqn.~\ref{intone2} \cite{wang} and Hamann employed Eqn.~\ref{intone1} \cite{hamann} 
as the starting points for equal--time Stratonovich--Hubbard decoupling schemes.
Note that both these expressions imply an Ising--like S--H field governing the spin degrees
of freedom, while the third expression implies an xy--like field. All three expressions result in
actions which break the spin--rotational symmetry of the theory, and the symmetry is only
restored by the integration over field configurations. As long as all configurations are 
included there is no difficulty, but problems arise if approximations are introduced for any
system beyond a single impurity. This was realized by Evenson et al. \cite{evenson} in large--$U$
studies of the two--site Hubbard model within a static approximation: The effective spin--spin
coupling assumes an Ising, rather than Heisenberg, form when Eqn.~\ref{intone2} is used as
a starting point. It would clearly be desirable to introduce a
symmetrical Heisenberg decoupling, but this is difficult for the following reason: No matter how
the spin quantization axis is chosen, the three
equivalent components of the spin--spin interaction are divided up between the
two particle--hole channels (longitudinal and transverse) of the electron vertex, which ``overlap''
completely at equal time. A Heisenberg
decoupling can be introduced artificially if, for example, the interaction is written as an equal--weight
linear combination of Eqn.~\ref{intone1} for the three choices of moment operator, i.e.,
\beq
U\widehat n_{i\uparrow}\widehat n_{i\downarrow}\ =\ \fourth U\widehat n_i^2 -\twelfth U
\opvecm_i\cdot\opvecm_i\ .
\label{intone4}
\eeq
This procedure introduces spurious factors of $\third$ in the saddle--point equations when a
direction is chosen for the broken--symmetry spin axis, however, and
so offers an unsatisfactory solution.

We may restate the problem as follows: A conventional equal--time S--H transformation, followed 
by a static approximation within the resulting theory, violates the crossing symmetry of the Hubbard
interaction vertex and prevents a spin--rotation--invariant treatment of fluctuations.
To insure an equivalent static treatment of all three spin components, we propose the
introduction of a zero--frequency, rather than equal--time, S--H field. Such a decoupling scheme is
possible because the static components of the direct and exchange channels in the Hubbard model 
have essentially zero overlap at low temperatures. This allows the simultaneous treatment of
interactions in both crossed channels and leads to a unique prescription for the decoupling scheme
independent of the expression employed for the interaction in Eqns.~\ref{intone1}-\ref{intone4}. The
same procedure can be applied in more general situations when it is necessary to decouple multiple
interactions in crossed channels.

It is possible to carry out this transformation directly within an operator--based approach,
but the manipulations are more transparent with anticommuting c--numbers, and we follow the
latter approach. We consider a discretized representation of the partition function
\beq
Z\ =\ {\rm Tr}\,e^{-\beta \widehat H}\ ,
\label{partition}
\eeq
where
\beq 
\widehat H\ =\ -t\!\sum_{i\delta\sigma}\,c^{\dag}_{i+\delta,\,\sigma}c^{\vphantom{\dag}}
_{i\sigma}\ +\ U\sum_i\,c^{\dag}_{i\uparrow}c^{\vphantom{\dag}}_{i\uparrow}c^{\dag}_{i\downarrow}
c^{\vphantom{\dag}}_{i\downarrow}\ -\ \mu\sum_{i\sigma}\,c^{\dag}_{\i\sigma}c^{\vphantom{\dag}}_
{i\sigma}\ .
\label{hamone}
\eeq
Let
\beqa
\tau_{\ell} &=& \ell\Delta\tau \nonumber \\
\Delta\tau &=& \beta/L\ ,
\label{tau}
\eeqa
with $L$ the number of time slices. We write the anticommuting c--numbers for time $\tau_{\ell}$,
site $i$, and spin $\sigma$ as $\cbar^{i\sigma}_{\ell}$, $c^{i\sigma}_{\ell}$. It is convenient
to use the frequency--transformed variables $\cbar^{i\sigma}_n$, $c^{i\sigma}_n$, where
\beqa
\cbar^{i\sigma}_{\ell}\ &=&\ \sum_{n=-L/2}^{L/2-1}\,e^{i\omega_n\tau_{\ell}}\,\,
  \cbar^{i\sigma}_n \nonumber \\
 c^{i\sigma}_{\ell}\ &=&\ \sum_{n=-L/2}^{L/2-1}\,e^{-i\omega_n\tau_{\ell}}\,
  c^{i\sigma}_n 
\label{fourier}
\eeqa
with
\beq
\omega_n = (2n+1)\pi T
\label{matsubara}
\eeq
for integer $n$. In order to keep track of discretization effects, we define
\beq
i\omega_n^{\Delta}\ \equiv\ { 1-e^{-i\omega_n\Delta\tau} \over \Delta\tau}\ ,
\label{discrete}
\eeq
which approaches $i\omega_n$ for $\Delta\tau\rightarrow 0$.

The discretized fermion action is
\beqa
S\ &=&\ -\beta\sum_{in\sigma}\,(i\omega_n^{\Delta}+\mu)\,\cbar^{i\sigma}_n c^{i\sigma}_n
  \ +\ \beta\sum_{i\delta n\sigma}\,(-t)\,\cbar^{i+\delta,\,\sigma}_n c^{i\sigma}_n \nonumber \\
  &\phantom{=}& \qquad\qquad +\ \beta U\!\sum_{imnn'}\cbar^{i\uparrow}_{n+m}
  \cbar^{i\downarrow}_{-n-1} c^{i\downarrow}_{-n'-1}c^{i\uparrow}_{n'+m}\ .
\label{actionone}
\eeqa
\noindent
The interaction term is represented in Figure 1. (Note that the integer label $-n-1$
corresponds to frequency $\omega_{-n-1}=-\omega_n$.) 
\noindent Equivalently one may write
\beq
S\ =\ S_0 + S_{\rm int}\ ,
\label{actiontwo}
\eeq
with
\beq
S_0\ =\ \beta\!\sum_{ijn\sigma}\, \cbar^{i\sigma}_n\Bigl[\,-i\omega_n^{\Delta}{\bf 1}
  + ({\bf H_0}-\mu{\bf 1})\,\Bigr]_{ij}c^{j\sigma}_n\ ,
\label{actionthree}
\eeq
where the matrix in brackets has spatial indices, and
\beqa
{\bf 1}_{ij}\ &=&\ \delta_{ij} \nonumber \\
({\bf H_0})_{ij}\ &=&\ \cases{-t, & $i$ and $j$ near--neighbors \cr
                             0, & else. \cr} 
\label{mats}
\eeqa

Now note that the interaction can be rewritten in two equivalent ways to emphasize the
particle--hole channels (see Figures 2a--b):
\beqa
S_{\rm int}\ &=&\ -\beta U\sum_{imnn'} \cbar^{i\uparrow}_{n+m}c^{i\downarrow}_n
  \cbar^{i\downarrow}_{n'}c^{i\uparrow}_{n'+m} \nonumber \\
            &=&\ \phantom{-}\beta U\sum_{imnn'} \cbar^{i\uparrow}_{n+m}
  c^{i\uparrow}_n\cbar^{i\downarrow}_{n'}c^{i\downarrow}_{n'+m}\ .
\label{crossing}
\eeqa
\noindent
At this stage we restrict attention to terms with $m=0$, i.e., the static component of the
particle--hole interactions. For $m=0$ the two sums in Eqn.~\ref{crossing} are essentially
independent: The only terms which appear in both sums are restricted to $n'=n$. Therefore 
the static particle--hole interaction may be written 
\beq
S^{m=0}_{\rm int}\ =\ -\beta U\sum_i\,\Bigl[\,\sum_n\,\cbar^{i\uparrow}_n
  c^{i\downarrow}_n\,\Bigr]\,\Bigl[\,\sum_{n'}\,\cbar^{i\downarrow}_{n'}c^{i\uparrow}_{n'}\,
  \Bigr]\ +\ \beta U\sum_i\,\Bigl[\,\sum_n\,\cbar^{i\uparrow}_nc^{i\uparrow}_n\,
  \Bigr]\,\Bigl[\,\sum_{n'}\,\cbar^{i\downarrow}_{n'}c^{i\downarrow}_{n'}\,\Bigr]
\label{actionfour}
\eeq
where we have dropped from the double sums an $n'=n$ correction term, which becomes negligible at low 
temperatures \cite{correction}.

Now all terms in Eqn.~\ref{actionfour} may be reduced to sums of perfect squares 
in preparation for the introduction of S--H fields. First note that
\beqa 
\sum_n\,\cbar^{i\uparrow}_nc^{i\downarrow}_n\ &=&\ \half\sum_{n\alpha\alpha'}\,
  \cbar^{i\alpha}_n\,(\sigma_x+i\sigma_y)_{\alpha\alpha'}\,c^{i\alpha'}_n \nonumber \\
\sum_n\,\cbar^{i\downarrow}_nc^{i\uparrow}_n\ &=&\ \half\sum_{n\alpha\alpha'}\,
  \cbar^{i\alpha}_n\,(\sigma_x-i\sigma_y)_{\alpha\alpha'}\,c^{i\alpha'}_n \ .
\label{sumsone}
\eeqa
Using these identifications one has
\beq
\Bigl[\,\sum_n\,\cbar^{i\uparrow}_nc^{i\downarrow}_n\,\Bigr]\,\Bigl[\,\sum_{n'}\,
  \cbar^{i\downarrow}_{n'}c^{i\uparrow}_{n'}\,\Bigr]\ =\ \fourth\biggl[\,(m^i_{x0})^2 +
  (m^i_{y0})^2\,\biggr]\ ,
\label{sumstwo}
\eeq
where
\beqa
  m^i_{x0}\ &=&\ \sum_{n\alpha\alpha'}\,\cbar^{i\alpha}_n\,(\sigma_x)_{\alpha\alpha'}\,
  c^{i\alpha'}_n \nonumber \\
  m^i_{y0}\ &=&\ \sum_{n\alpha\alpha'}\,\cbar^{i\alpha}_n\,(\sigma_y)_{\alpha\alpha'}\,
  c^{i\alpha'}_n \ .
\label{sumsthree}
\eeqa
\noindent Likewise,
\beqa
\sum_n\,\cbar^{i\uparrow}_nc^{i\uparrow}_n\ &=&\ \half(n^i_0+m^i_{z0}) \nonumber \\
\sum_n\,\cbar^{i\downarrow}_nc^{i\downarrow}_n\ &=&\ \half(n^i_0-m^i_{z0}) \ ,
\label{sumsfour}
\eeqa
where
\beqa
n^i_0\ &=&\ \sum_n\,(\cbar^{i\uparrow}_nc^{i\uparrow}_n+\cbar^{i\downarrow}_nc^{i\downarrow}_n) 
  \nonumber \\
m^i_{z0}\ &=&\ \sum_{n\alpha\alpha'}\,\cbar^{i\alpha}_n\,(\sigma_z)_{\alpha\alpha'}
  \,c^{i\alpha'}_n \ .
\label{sumsfive}
\eeqa
In this case one identifies
\beq
\Bigl[\,\sum_n\,\cbar^{i\uparrow}_nc^{i\uparrow}_n\,\Bigr]\,\Bigl[\,\sum_{n'}\,
  \cbar^{i\downarrow}_{n'}c^{i\downarrow}_{n'}\,\Bigr]\ =\ \fourth\biggl[\,(n^i_0)^2 -
  (m^i_{z0})^2\,\biggr]\ .
\label{sumssix}
\eeq
Thus
\beq
S^{m=0}_{\rm int}\ =\ \fourth\beta U\sum_i\,(n^i_0)^2\ -
  \ \fourth\beta U\sum_i\,(\vecm^i_0\cdot\vecm^i_0)^2\ .
\label{actionsix}
\eeq

The repulsive charge interaction is retained for treatment by perturbation theory (i.e.,
saddle--point plus fluctuations). The attractive spin interaction in Eqn.~\ref{actionsix} may be decoupled using an
independent static S--H field at each site $i$:
\beqa
 &\phantom{=}&\exp\Bigl[\,-S^{m=0}_{\rm spin}\,\Bigr]\ =
  \ \exp\Bigl[\,\fourth\beta U\sum_i
  \,\vecm^i_0\cdot\vecm^i_0\,\Bigr] \nonumber \\
  &\phantom{=}&\qquad = \int\,\prod_i\Bigl( {\beta\over\pi}\Bigr)^{3/2}
  \,d\xi^i_x\,d\xi^i_y\,d\xi^i_z\,\,
  \exp\biggl[\,-\beta\sum_i\,\xii\cdot\xii
  \ +\ \beta\sqrt{U}\,\sum_i\,\xii \cdot\vecm^i_0\,\biggr]\ . \nonumber\\
  &\phantom{=}&
\label{stratonovich}
\eeqa

The S--H fields $\xii$ may be interpreted as classical magnetic fields
which couple to the local electron spin polarization. The Hubbard action may be rewritten as
\beq
S\ =\ \beta\sum_i\,|\xi^i|^2\ +\ \widetilde S_0\ +\ S^{m=0}_{\rm charge}\ +\ S^{m
  \neq 0}_{\rm int}\ ,
\label{actionseven}
\eeq
where
\beq
\widetilde S_0\ =\ \beta\sum_{ijn\alpha\alpha'}\cbar^{i\alpha}_n\,\Bigl\{\,-i\omega^{\Delta}
  _n\delta_{\alpha\alpha'}\delta_{ij}\ +\ \biggl[\,({\bf H_0})_{ij}\delta_{\alpha\alpha'}\,
  -\,\sqrt{U}\,\xii\cdot\vec\sigma_{\alpha\alpha'}\,\delta_{ij}\,-\,\mu
  \delta_{ij}\delta_{\alpha\alpha'}\,\biggr]\,\Bigr\}\,c^{j\alpha'}_n
\label{actioneight}
\eeq
and
\beq
S^{m=0}_{\rm charge}\ =\ \fourth\beta U\sum_i\,(n^i_0)^2\ .
\label{actionnine}
\eeq

The interaction contributions to the action for $m\neq 0$ take precisely the form they would
in a conventional perturbative treatment, {\sl except} that all vertices related to the
zero--frequency spin vertex by crossing are absent. Note that because we have decoupled only
the $m=0$ spin components, we have left open the possibility of higher order perturbative
expansions about a static approximation. 

We consider now the
static charge interaction $S^{m=0}_{\rm charge}$ in Eqn.~\ref{actionseven}. This term
could be decoupled in parallel with $S^{m=0}_{\rm spin}$ by introducing a
real--valued S--H field $\rho^i$. Since the charge interaction is repulsive, 
however, the
fermion bilinear coupling to $\rho^i$ would be pure imaginary and unsuitable for Monte Carlo
simulation. It is straightforward to show that a saddle--point approximation
for the $\rho^i$ fields in the presence of a fixed $\xii$ configuration yields the same
result as a spatially non--uniform Hartree--Fock. For this reason we do
not introduce $\rho^i$ fields, but reserve $S^{m=0}_{\rm charge}$ for treatment by perturbation
theory (i.e., Hartree--Fock at lowest order).

Within such a Hartree--Fock approximation for fixed $\xii$,
\beq
S^{m=0}_{\rm charge}\ \rightarrow\ \half\beta U\sum_{in\alpha}\,\langle n^i\rangle_{\xi}\,
  \cbar^{i\alpha}_nc^{i\alpha}_n\ -\ \fourth\beta U\sum_i\,\langle n^i\rangle_{\xi}^2\ ,
\label{actionten}
\eeq
where $\langle n^i\rangle_{\xi}$ is the mean occupancy at site $i$.
It is essential to retain the ``constant'' term, since it depends on the local field configuration
$\{\xii\}$. From this point we use the notation $\langle A\rangle_{\xi}$ to denote the
expectation value of the fermion observable $A$ in the presence of a fixed field configuration.

The static $\vec\xi$--dependent action assumes the form
\beq
\overline S(\xi)\ =\ \overline S_0(\xi)\ +\ \overline S_{\rm f}(\xi,\,\cbar,\,c) \ ,
\label{actioneleven}
\eeq
where
\beqa
\overline S_0(\xi)\ &=&\ \beta\sum_i\,|\xi^i|^2\ -\ \fourth\beta U\sum_i\,
  \langle n^i\rangle_{\xi}^2 \nonumber \\
\overline S_{\rm f}(\xi,\,\cbar,\,c)\ &=&\ \beta\sum_{ijn\alpha\alpha'}\cbar^{i\alpha}_n\Bigl\{
  \,-i\omega^{\Delta}_n\delta_{ij}\delta_{\alpha\alpha'}\ +\ \bigl[\,({\bf H_0})_{ij}\delta_
  {\alpha\alpha'}\,-\,\sqrt{U}\xii\cdot\vec\sigma_{\alpha\alpha'}\delta_{ij} \nonumber\\
  &\phantom{=}&\qquad\qquad +
  \ \bigl(\half U\langle n^i\rangle_{\xi}-\mu\bigr)\delta_{ij}\delta_{\alpha\alpha'}\,\bigr]\,
  \Bigr\}c^{j\alpha'}_n\ .
\label{actiontwelve}
\eeqa
The fermion contribution to the action may now be integrated out exactly, since it takes a one--body
form (albeit a form non--diagonal in spin):
\beq
\int\,D\cbar\,Dc\, \exp[-\overline S_{\rm f}]\ =\ \prod_r\,\Bigl[1+e^{-\beta\varepsilon_r(\xi)}\Bigr]\ ,
\label{fermion}
\eeq
where the $\varepsilon_r$ are eigenvalues of the one--body Hamiltonian
\beq
H(\xi)\ =\ -t\sum_{i\delta\alpha}\,c^{\dag}_{i+\delta,\,\alpha}c^{\vphantom{\dag}}_{i\alpha}
  \ -\ \sqrt{U}\sum_{i\alpha\alpha'}\,\xii\cdot c^{\dag}_{i\alpha}\vec\sigma_{\alpha\alpha'}
  c^{\vphantom{\dag}}_{i\alpha'}\ +\ \sum_i\,\biggl[\,\half U\langle n^i\rangle_{\xi}-\mu\,\biggr]
  \,c^{\dag}_{i\alpha}c^{\vphantom{\dag}}_{i\alpha}\ . 
\label{hamtwo}
\eeq
Thus the final expression for the effective action after the fermion integration is
\beq
S_{\rm eff}(\xi)\ =\ \overline S_0(\xi)\ -\ \sum_r\,\log\,\bigl[1+e^{-\beta\varepsilon_r(\xi)}\bigr]
  \ .
\label{newaction}
\eeq

Note that for fixed $\mu$ and $\xii$ a Hartree--Fock iteration is required to adjust the 
non--uniform charge field $\langle n^i\rangle$ to satisfy the condition
\beqa
\langle n^i\rangle\ &=&\ \sum_r\,f(\varepsilon_r)\sum_{\sigma}\,|\langle r|i\sigma\rangle|^2
  \nonumber \\
f(\varepsilon_r)\ &=&\ {1\over 1+e^{\beta\varepsilon_r} }\ .
\label{scfone}
\eeqa
In this equation $|r\rangle$ is the one--body eigenvector corresponding to eigenvalue 
$\varepsilon_r(\xi)$;
and the $|i\sigma\rangle$ constitute a complete set of basis vectors corresponding to spin $S_z=\sigma$ 
on site $i$. Note that for an $\nsites\times\nsites$ real--space lattice, the size of the eigenvector space is $2\nsites$.

The form of the charge--interaction Hartree--Fock employed here provides a template for the
family of Baym--Kadanoff \cite{baym} (or FLEX \cite{bs}) generalizations of the present approach discussed 
in Section 6. The charge field is here viewed as a nonsingular ``slave'' to the spin field: It is fixed to its
saddle--point, which changes when the $\vec\xi$ configuration changes.

No further approximation to the action $S_{\rm eff}(\xi)$ is now possible without
explicitly breaking spin rotational and translational invariance. In conventional Hartree--Fock
treatments it is assumed that the static field $\xii$ assumes a rigid configuration.
At half--filling the mean--field ansatz is just 
\beqa
\xii\ &=&\ \vec\xi\,e^{iQ\cdot R_i}\,\qquad Q\,=\,(\pi,\,\pi) \nonumber\\
\langle n^i\rangle\ &=&\ 1\ . 
\label{ansatz}
\eeqa
A saddle--point evaluation of $S_{\rm eff}$ within this restricted field space reproduces the
conventional Hartree--Fock equations for a commensurate antiferromagnet. Furthermore inhomogeneous
Hartree--Fock solutions \cite{zaanen} for $\langle n\rangle \neq 1$ follow by assuming a spin--space form
(collinear, spiral, etc.) for the order parameter and solving the resulting saddle--point equations.

We propose instead to evaluate electron correlation functions using $S_{\rm eff}$
as the action in a classical field Monte Carlo. The Monte Carlo is classical since all 
finite--frequency contributions to the action have been set aside for treatment by perturbation
theory: Such contributions, when included, alter the form of the static field action, but are not
themselves statistically sampled. By this means it is possible to insure that higher--order
approximations have a positive--definite weighting factor $\exp(-S_{\rm eff})$, avoiding the sign
problem inherent in a full quantum Monte Carlo.

It is important to note that this approach has been explicitly designed to treat the low--temperature
regime in which local moments exist, i.e., in which the most probable configurations of
$\xii$ have non--zero amplitude (at least for some sites $i$). We do not address in this
paper the problem of local moment formation per se. It is clear that at high temperatures the
``free'' $\xii$ fluctuations have variance $\langle\,\xii\cdot\xii\,\rangle\,
\propto\, T$, and it is not appropriate to use $S_{\rm eff}(\xi)$ as a starting point for
perturbation theory in this regime. Instead our purpose is to construct an approximation
scheme which builds in the presence of disordered local moments at temperatures satisfying
\beq 
4t^2/U\ \leq\ T\ \ll\ U
\label{condition1}
\eeq
in three dimensions and
\beq 
T\ \ll\ U
\label{condition2}
\eeq
in two dimensions.

It is also important to note that this approximation omits quantum effects associated with
``tunneling'' of the order parameter fields, i.e., possible $\tau$--dependent instantons.
As a result the Kondo effect is outside the present description. We restrict our attention here
to the implications of a classical (i.e., $m=0$) order parameter description
and do not discuss additional quantum effects. 

We next examine the behavior of $S_{\rm eff}$ in the
large--$U$ limit, then comment on the action's relationship to other work in the literature. In the large--$U$
limit the Hubbard model may be mapped to a quantum spin--$\half$
Heisenberg model with near--neighbor antiferromagnetic exchange integral $J=4t^2/U$. We show below that
in this limit the approximate action
$S_{\rm eff}$ describes a classical Heisenberg model with exchange integral
$J=4t^2/U$. This insures that in a two--dimensional system the approximation 
obeys the Mermin--Wagner Theorem, i.e., 
infinite--range spin ordering is absent down to $T=0$. However this approach does not
include the quantum fluctuations which renormalize the zero--temperature moment and reduce the
finite--temperature correlation length. It is essential to include finite--frequency fluctuations
in order to address these points. Nevertheless the static action $S_{\rm eff}$ achieves the goal
of reproducing the exact classical critical behavior 
within a well--defined zeroth--order approximation.

To demonstrate the mapping to the classical Heisenberg model, consider first the 
local one--site action ($t=0$) in the large--$U$ limit for $\langle n\rangle=1$:
\beq
\overline S_{\rm eff}^{\rm one-site}\ =\ \beta|\xi|^2\ -\ \sum_r\,\log\,\bigl[1+e^{-\beta
  \varepsilon_r(\xi)} \bigr]\ .
\label{onesite}
\eeq
In this case there are only two eigenvalues $\varepsilon_r$. To find $\varepsilon_r$, choose
the spin quantization axis along the direction of $\vec\xi$. Then the field--dependent term in
$H(\xi)$ is just
\beq
-\sqrt{U}\xi\,(c^{\dag}_{i\uparrow}c^{\vphantom{\dag}}_{i\uparrow}-c^{\dag}_{i\downarrow}
  c^{\vphantom{\dag}}_{i\downarrow})\ ,
\label{intern}
\eeq
with eigenvalues
\beq
\varepsilon_{\alpha}\ =\ -\sqrt{U}\xi\alpha,\qquad \alpha=\pm 1\ .
\label{eigs}
\eeq
Minimization of the action with respect to $\xi$ gives
\beq
\xi\ =\ \sqrt{U}\sum_{\alpha}\,{\alpha\,e^{\beta\alpha\sqrt{U}\xi}\over 1+e^{\beta\alpha\sqrt{U}\xi} }
  \ ,
\label{minimiz}
\eeq
i.e., 
\beq
\xi\ =\ \half\sqrt{U}\,\langle m_z\rangle\ ,
\label{solution}
\eeq
since
\beq
\langle m_z\rangle\ =\ \sum_{\alpha=\pm 1}\,\alpha\,{1\over 1+e^{-\beta\alpha\sqrt{U}\xi} }\ .
\label{mzz}
\eeq
For large $U$ the solution has
\beqa
\langle m_z\rangle\ &\rightarrow&\ 1 \nonumber \\
\xi\ &\rightarrow&\ \sqrt{U}/2\ .
\label{limits}
\eeqa
As expected, the one--site gap is just
\beq
\varepsilon_{\downarrow}-\varepsilon_{\uparrow}\ =\ \half U-(-\half U)\ =\ U\ .
\label{gapp}
\eeq
Note that this result has been obtained without breaking the rotational invariance of
the one--site Hamiltonian, since the direction of $\vec\xi$ is arbitrary, and the
Boltzmann weight must be integrated over $\vec\xi$.

Thus to zeroth order in $t/U$, the half--filled system consists of fluctuating fields
$\xii$ of fixed length $\sqrt{U}/2$ and uncorrelated directions. The first non--zero terms
in perturbation theory enter at $O(t/U)^2$, and describe the process of one electron hopping to
a neighboring site, then back to its starting point. Physically the hopping terms should induce
correlations in the field directions at neighboring sites to lower the kinetic energy. Let the
field directions at neighboring sites $i$ and $j$ be $\widehat n^i$ and $\widehat n^j$. Suppose
also that $\widehat n^i$ may be obtained by rotating $\widehat z$ in a right--handed sense by angle
$\alpha$ about axis $\widehat \alpha\perp\widehat z$. Then the eigenstates with spin projection
$\pm\half$ along $\widehat n^i$ can be written
\beqa
|\widehat n^i,\,\uparrow\rangle\ &=&\ \cos{\alpha\over 2}\,|\widehat z,\,\uparrow\rangle
  \ +\ (-i\alpha_x+\alpha_y)\,\sin {\alpha\over 2}\,|\widehat z,\,\downarrow\rangle \nonumber\\
|\widehat n^i,\,\downarrow\rangle\ &=&\ (-i\alpha_x-\alpha_y)\,\sin {\alpha\over 2}\,
  |\widehat z,\,\uparrow\rangle \ +\ \cos{\alpha\over 2}\,|\widehat z,\,\downarrow\rangle\ ,
\label{transform}
\eeqa
where
\beq
\alpha_x^2+\alpha_y^2\ =\ 1\ .
\label{sumsquares}
\eeq
The one--site solution gives
\beq
|\xii|\ =\ \half\sqrt{U}\ =\ \half\sqrt{U}\,\langle\,\widehat n^i\cdot\vecm^i
 \,\rangle\ ,
\label{meanfield}
\eeq
so that $|\widehat n^i,\,\uparrow\rangle$ will be occupied and $|\widehat n^i,\,\downarrow\rangle$
will be empty (and higher in energy by $U$). Now apply second--order perturbation theory to
the electron ground state for the frozen field configuration with direction vectors $\{\widehat
n^i\}$. The matrix element for hopping from
\beq 
|\widehat n^i,\,\uparrow\rangle\,|\widehat n^j,\,\uparrow\rangle\ \rightarrow
  \ |\widehat n^j,\,\downarrow\rangle\,|\widehat n^j,\,\uparrow\rangle 
\label{hoppingsites}
\eeq
is
\beq
(-t)\,\langle \widehat n^j,\,\downarrow|\,\widehat n^i,\,\uparrow\rangle\ =\ (-t)
  \,(-i\theta_x-\theta_y)^*\,\sin{\theta\over 2}\ ,
\label{hoppingmatrix}
\eeq
where $\widehat n^j$ is obtained from $\widehat n^i$ by a right--handed rotation by $\theta$
about axis $\widehat\theta\perp\widehat n^i$. The energy lowering of the ground state due to the
virtual hop from $i\rightarrow j$ and back is then
\beq
  {|\,(-t)(-i\theta_x-\theta_y)\,\sin(\theta/2)\,|^2\over -U}\ =\ -{t^2\over U}\,\sin^2{\theta\over 2}
  \ ,
\label{lowering}
\eeq
since
\beq \theta_x^2+\theta_y^2\ =\ 1\ .
\label{sumup}
\eeq
The hopping from $j\rightarrow i$ supplies the same energy, so the total perturbation
energy lowering in this particular field configuration is
\beqa
-{2t^2\over U}\sum_{\langle ij\rangle}\,\sin^2(\theta_{ij}/2)\ &=&\ {t^2\over U}\sum_{\langle ij
\rangle}\,(\cos\theta_{ij}-1) \nonumber\\
  &=&\ {4t^2\over U}\sum_{\langle ij\rangle}\,\biggl[\,\fourth\widehat n^i\cdot\widehat n^j
  \,-\,\fourth\biggr] \nonumber \\
  &=&\ {4t^2\over U}\sum_{\langle ij\rangle}\,\biggl[\,\langle \vec S^i\rangle_{\xi}\cdot
  \langle\vec S^j\rangle_{\xi}\,-\,\fourth\,\biggr]\ , 
\label{heisen}
\eeqa
where $\langle\vec S^i\rangle_{\xi}$ is the electron spin expectation value in configuration 
$\{\xii\}$ for $t=0$. Note finally that the effective action at half--filling reduces to
\beq 
S_{\rm eff}(\xi)\ =\ \beta\sum_i\,|\xi^i|^2\ -\ \log\,\prod_r\,\bigl[1+e^{-\beta\varepsilon_r
  (\xi)}\bigr]\ .
\label{actionthirteen}
\eeq
In the large--$U$ limit
\beq
\prod_r\,\bigl[1+e^{-\beta\varepsilon_r(\xi)}\bigr]\ \simeq\ \exp\bigl[-\beta E_0(\xi)\bigr]\ ,
\label{prodd}
\eeq
with $E_0$ the perturbed ground state energy, since all other occupancies result in an energy
higher by at least $U$. Since $\beta\sum_i\,|\xi^i|^2$ is a constant in this limit, the part
of the action dependent on the fluctuating fields is precisely $\beta H_{\rm eff}$, with
\beq 
H_{\rm eff}\ =\ {4t^2\over U}\sum_{\langle ij\rangle}\,(\,\fourth \widehat n^i\cdot\widehat n^j
  \,-\,\fourth)\ ,
\label{hamthree}
\eeq
establishing the mapping to the classical Heisenberg model.

The behavior of the action away from half--filling in the large--$U$ limit is not so easy to
deduce, since the Hartree--Fock correction from the static charge interaction is crucial. If 
one assumes $N_{\rm el}$ sites are singly occupied and $\nsites-N_{\rm el}$ sites are empty,
the one--site saddle--point equation implies $|\xi^i|\,=\,\sqrt{U}/2$ on the singly
occupied sites and $|\xi^i|\,=\,0$ on the empty sites. A degeneracy then arises, since the
low--lying one--electron states on the singly occupied sites pick up a Hartree--Fock density
contribution $(U/2)\langle n^i\rangle=U/2$, while the states on the empty sites have no
Hartree--Fock correction. The two sorts of sites then have degenerate low--lying states in the
absence of $t$: The occupied state on each singly occupied site has energy $-\half U+\half U=0$,
and the two states on each unoccupied site also have energy 0. As expected, the hopping becomes
crucial for determining $\langle n^i\rangle$ and the favored $\xii$ configurations. Thus in the
doped case there are important amplitude, as well as orientation, fluctuations in the
S--H field configurations arising from $S_{\rm eff}$.

We next comment briefly on the relationship of our approximation to other work in the
literature. We have argued in Section 1 that perturbation theory in $U$ fails to resum
properly at low temperatures in the doped Mott insulator phase due to a fundamental change in
the system's saddle--point structure. Nevertheless a number of authors \cite{moukouri}--\cite{vilk} have 
observed that gaps or pseudogaps in the one--electron density of states for the positive--$U$ and 
negative--$U$ Hubbard models may be obtained within approximations which resemble 
perturbation theory about a non--interacting ground state. Such approximations 
have in common the feature that the one--electron self--energy $\Sigma$ may be written in the form
\beq \Sigma(k,\,i\omega_n)\ =\ {T\over\nsites}\sum_{Qm}\, V(Q,\,i\Omega_m)G_0(k-Q,\,i(\omega_n-
  \Omega_m))\ ,
\label{perturb}
\eeq
where $V$ is a momentum-- and frequency--dependent effective potential and $G_0$ is a bare
electron propagator. Crucial points to note are that (i)~the electron propagator is left
bare; and (ii)~higher order perturbation theory in $V$
leads to a deterioration, rather than an improvement, in the approximation for the 
one--electron density of states. We argue that the reasons for this behavior are as follows:
The local moments which underlie the Mott--Hubbard gap generate anomalous diagrams in 
perturbation theory, i.e., terms in the one--electron self--energy off--diagonal in momentum and
spin. At zero temperature and half filling the moments
have true long--range staggered order, so that a momentum--space description becomes possible. If the
static off--diagonal self--energy is denoted $\Delta_{\widetilde Q}$, with $\widetilde Q=(\pi,\,\pi)$, 
then the effective diagonal self--energy becomes
\beq
\Sigma(k,\,i\omega_n)\ =\ \Delta^2_{\widetilde Q}\,G_0(k-\widetilde{Q})\ ,
\label{afmhartree}
\eeq
generating a Hartree--Fock antiferromagnetic gap of $2\Delta_{\widetilde Q}$ in the density of
states. In this case, as is well known, $G_0$ enters Eqn.~\ref{afmhartree} rather than $G$ to
avoid double counting. Thus approximations
of the form in Eqn.~\ref{perturb} may be viewed as generalizations of Eqn.~\ref{afmhartree} 
which ``smear'' the off--diagonal self--energy $\Delta$ over a range in momentum-- and frequency--space
\cite{lee}. In this sense these approximations are quite different from perturbation theory in $V$, and they
do not provide support for the convergence of perturbation theory in the doped Mott insulator. 
We would argue that such approximations model only roughly the physics of this
system \cite{rough}.  In the doped phase the order parameter field has strong non--Gaussian fluctuations in both 
amplitude and orientation, and these are coupled to the charge degrees of freedom. Furthermore, as
discussed in reference~\cite{millis}, this general approach has difficulties addressing rare large
amplitude fluctuations that play an essential role.

We comment also on the relationship of the current approach to the phenomenological spin--fermion
(SF) model. In recent years a number of perturbative studies \cite{altshuler,chubukov,schmalian} have been 
carried out for this model, which describes a system of electrons 
coupled to a Gaussian--distributed vector spin field ${\bf S}(Q,\,i\Omega)$. The microscopic action $S_{\rm eff}$
differs from $S_{\rm SF}$ in two crucial ways: (i)~$S_{\rm SF}$ omits coupling to charge
degrees of freedom, while $S_{\rm eff}$ describes a highly nonlinear coupling between spin and
charge. The form of this coupling is dictated by the Hubbard interaction itself,
and its omission prevents a proper description of the doped Mott insulator phase. 
(ii)~The finite--frequency corrections to the actions differ significantly, since we have introduced only
a static S--H field to circumvent the problems associated with an
equal--time decoupling. Finally, our point on the non--Gaussian nature of fluctuations in $S_{\rm eff}$
applies to $S_{\rm SF}$ as well, i.e., in the interesting parameter regime for this model we expect
a conventional perturbative treatment to be impossible due to the formation of local moments.

\bigskip\bigskip
\section{Evaluation of electron correlation functions}

Electron correlation functions for the action $\overline S$ in Eqn.~\ref{actioneleven}
have been calculated using the Metropolis
algorithm for statistical sampling of $\xii$ configurations.The Monte Carlo algorithm
adopted relies on single--site updating and resembles a hybrid of Glauber and Kawasaki
dynamics \cite{newman} for the Ising or classical Heisenberg models. The algorithm is described more precisely
in Section 4.

Electron correlation functions have the same general form as in full quantum Monte Carlo
calculations with continuous--valued S--H fields. Let $A(\cbar,\,c)$ be an
electron observable written in anticommuting c--number notation, and let
\beq
D\xi\ \equiv\ \int\,\prod_i\,\Bigl( {\beta\over\pi}\Bigr)^{3/2}\,d\xi^i_x\,d\xi^i_y\,d\xi^i_z\ .
\label{measure}
\eeq
Then
\beqa
\langle A\rangle\ &=&\ { \int\,D\xi\,\exp[-\overline S_0(\xi)]\,\int\,D\cbar\,Dc\,\, A(\cbar,\,c)
  \,\exp[-\overline S_{\rm f}(\xi,\,\cbar,\,c)] \over \int\,D\xi\,D\cbar\,Dc\,\,\exp[-\overline S_0(\xi)]
  \,\exp[-\overline S_{\rm f}(\xi,\,\cbar,\,c)]} \nonumber\\
 &=&\ {1\over Z}\int\,D\xi\,\exp[-\overline S_0(\xi)]\,\Bigl[\, {\int\,D\cbar\,Dc\,\,A(\cbar,\,c)
  \,\exp[-\overline S_{\rm f}(\xi,\,\cbar,\,c)] \over \int\,D\cbar\,Dc\,\,\exp[-\overline S_{\rm f}
  (\xi,\,\cbar,\,c)] } \,\Bigr]\,Z_{\rm f}(\xi)\ , \nonumber\\
 &\phantom{=}& 
\label{expect}
\eeqa
where
\beqa
Z_{\rm f}(\xi)\ &=&\ \int\,D\cbar\,Dc\,\,\exp[-\overline S_{\rm f}(\xi,\,\cbar,\,c)] \nonumber \\
  &=&\ \prod_r\,\bigl[1+e^{-\beta\varepsilon_r(\xi)}\bigr]\ ,
\label{expect2}
\eeqa
and the last line in Eqn.~\ref{expect} follows by multiplying and dividing by $Z_{\rm f}$
within the integral. The expression in square brackets is just an electron thermal average
in the presence of the background $\xii$ configuration. Thus finally
\beqa
\langle A\rangle\ &=&\ {1\over Z}\int\,D\xi\,\exp[-S_{\rm eff}(\xi)]\,
  \langle A(\cbar,\,c)\rangle_{\xi} \nonumber \\
  &\equiv&\ \langle\,\langle A(\cbar,\,c)\rangle_{\xi}\,\rangle\ ,
\label{expect3}
\eeqa
where
\beqa
S_{\rm eff}(\xi)\ &=&\ \overline S_0(\xi)\ -\ \log Z_{\rm f}(\xi) \nonumber\\
  &=&\ \overline S_0(\xi)\ -\ \sum_r\,\log\,\bigl[1+e^{-\beta\varepsilon_r(\xi)}\bigr]\ .
\label{expect4}
\eeqa
The simple form for $S_{\rm eff}$ means that approximations which incorporate
finite--frequency fluctuations, i.e., which generalize $Z_{\rm f}(\xi)$, may in principle be 
developed to extend the approach worked out in detail here (see Section 6).

We derive below the form of several one-- and two--body electron correlation functions
$\langle A(\cbar,\,c)\rangle_{\xi}$ in the presence of the background fields. As stated
previously,
\beqa 
\langle n^i\rangle_{\xi}\ &=&\ \sum_{\alpha}\,G_{\alpha\alpha}(ii;\,\tau=0^-)_{\xi} \nonumber\\
  &=&\ \sum_{r\alpha}\,f(\varepsilon_r)\,|\langle r|i\alpha\rangle|^2\ ,
\label{density}
\eeqa
where
\beq
G_{\alpha\alpha'}(ij;\,\tau)_{\xi}\ =\ -\,\langle c^{i\alpha}(\tau)\cbar^{j\alpha'}(0)\rangle_{\xi} 
\label{gg}
\eeq
in terms of anticommuting c--numbers; and the $|r\rangle$ are eigenstates of the spin--dependent
Hamiltonian $H(\xi)$ in Eqn.~\ref{hamtwo}. Likewise, the spin polarization at site $i$ is
\beqa
\langle \vecm^i\rangle_{\xi}\ &=&\ \sum_{\alpha\alpha'}\,\langle \cbar^{i\alpha}\vec
  \sigma_{\alpha\alpha'}c^{i\alpha'}\,\rangle_{\xi} \nonumber\\
  &=&\ \sum_{\alpha\alpha'}\,\vec\sigma_{\alpha\alpha'}\,G_{\alpha'\alpha}(ii;\,\tau=0^-)_{\xi} 
  \nonumber\\
  &=&\ \sum_{r\alpha\alpha'}\,f(\varepsilon_r)\,\vec\sigma_{\alpha\alpha'}\,\langle r|i\alpha\rangle
  \,\langle i\alpha'|r\rangle\ ,
\label{polarization1}
\eeqa
in terms of the $S_z$--basis states $|i\alpha\rangle$. Writing out the vector components
explicitly gives
\beqa
\langle m^i_x\rangle_{\xi}\ &=&\ \sum_r\,f(\varepsilon_r)\,\Bigl[\,\langle i\downarrow|r\rangle\,
  \langle r|i\uparrow\rangle\ +\ \langle i\uparrow|r\rangle\,\langle r|i\downarrow\rangle\,\Bigr] 
  \nonumber\\
\langle m^i_y\rangle_{\xi}\ &=&\ \sum_r\,f(\varepsilon_r)\,\Bigl[\,-i\langle i\downarrow|r\rangle\,
  \langle r|i\uparrow\rangle\ +\ i\langle i\uparrow|r\rangle\,\langle r|i\downarrow\rangle\,\Bigr] 
  \nonumber\\
\langle m^i_z\rangle_{\xi}\ &=&\ \sum_r\,f(\varepsilon_r)\,\Bigl[\,|\langle i\uparrow|r\rangle|^2
  \ -\ |\langle i\downarrow|r\rangle|^2\,\Bigr]\ .
\label{polarization2}
\eeqa

Note that these last results generalize to give the one--body propagator
\beqa
G_{i\alpha,\,j\beta}(i\omega_n)_{\xi}\ &=&\ \int_0^{\beta}\,d\tau\,e^{i\omega_n\tau}\,
  \Bigl[\,-\langle c^{i\alpha}(\tau)\cbar^{j\beta}(0)\rangle_{\xi}\,\Bigr] \nonumber\\
  &=&\ -\beta\,\langle c^{i\alpha}_n\cbar^{j\beta}_n\rangle_{\xi} \nonumber\\
  &=&\ \sum_r\, {\langle i\alpha|r\rangle\,\langle r|j\beta\rangle \over i\omega_n-\varepsilon_r}\ .
\label{propagator}
\eeqa
The real--axis spectral density may be calculated directly by analytic continuation before the
field averaging is performed. Thus the one--particle spectral density for spin $\alpha$ at site
$i$ after field--averaging is
\beqa
N_{i\alpha}(\omega)\ &=&\ -{1\over\pi}\,\Bigl\langle\,{\rm Im}\,G_{\alpha\alpha}(ii;\,\omega+i0^+)
  _{\xi}\,\Bigr\rangle \nonumber\\
  &=&\ \Bigl\langle\,\sum_r\,|\langle i\alpha|r\rangle|^2\,\delta(\omega-\varepsilon_r)\,\Bigr
  \rangle\ .
\label{rhoa}
\eeqa
The total one--particle spectral density for spin $\alpha$ is
\beq
N_{\alpha}(\omega)\ =\ \sum_i\,N_{i\alpha}(\omega)\ . 
\label{rhotot}
\eeq
(Note that in a properly equilibrated calculation translational and spin--rotational invariance
are restored by field--averaging, and the quantity $N_{i\alpha}$ is site-- and 
spin-- independent.) The 
momentum--resolved spectral density $A_{k\alpha}(\omega)$ is
only slightly more difficult to calculate. Note again that translational invariance is only
restored by field--averaging, so that the propagator in a single field configuration is
not diagonal in momentum. After averaging we have
\beq
A_{k\alpha}(\omega)\ =\ \Bigl\langle\,\sum_r\,\delta(\omega-\varepsilon_r)\cdot
  {1\over\nsites}\sum_{ij}\,e^{-ik\cdot(R_i-R_j)}\,\langle i\alpha|r\rangle\,\langle r|j\alpha\rangle
  \,\Bigr\rangle
\eeq

All the spectral densities may be calculated conveniently by histogram binning, i.e., by
incrementing a bin centered on frequency $\omega_i$ whenever an eigenvalue $\varepsilon_r$
falls in the range $[\omega_i-(1/2)\Delta\omega,\,\omega_i+(1/2)\Delta\omega)$. In this way
spectral densities may be obtained which integrate exactly to unity within machine precision.

In some cases it may be desirable to compare imaginary--time propagators $G_{k\alpha}(\tau)$
with the results from a full quantum Monte Carlo simulation. The necessary expressions 
within the present approach are
\beq
G_{k\alpha}(\tau)\ =\ \cases{ \Bigl\langle\,-\sum_r\,e^{-\varepsilon_r\tau}\,[1-f(\varepsilon_r)]\,
  F^r_{k\alpha}\,\Bigr\rangle, & $\tau>0$ \cr \Bigl\langle\,\sum_r\,e^{-\varepsilon_r\tau}\,
  f(\varepsilon_r)\,F^r_{k\alpha}\,\Bigr\rangle, & $\tau<0$\cr}
\label{gtau}
\eeq
where
\beq
F^r_{k\alpha}\ =\ {1\over\nsites}\sum_{ij}\,e^{-ik\cdot(R_i-R_j)}\,\langle i\alpha|r\rangle
  \,\langle r|j\alpha\rangle\ .
\label{ffactor}
\eeq

A quantity of particular interest in the local moment regime is the average double occupancy
\beq
\langle D\rangle\ =\ {1\over\nsites}\sum_i\,\Bigl\langle\,\langle D^i\rangle_{\xi}\,\Bigr\rangle\ ,
\label{double1}
\eeq
where
\beq
\langle D^i\rangle_{\xi}\ =\ \langle n^{i\uparrow}n^{i\downarrow}\rangle_{\xi}\ .
\label{double2}
\eeq
In a general $\xii$ configuration with broken spin--rotational invariance the appropriate
expression for $\langle D^i\rangle_{\xi}$ is not $\langle n^{i\uparrow}\rangle_{\xi}\,\langle
  n^{i\downarrow}\rangle_{\xi}$. Instead
\beqa
\langle D^i\rangle_{\xi}\ &=&\ \lim_{\tau\rightarrow\tau'}\,\langle \cbar^{i\uparrow}(\tau)
  c^{i\uparrow}(\tau)\cbar^{i\downarrow}(\tau')c^{i\downarrow}(\tau')\rangle_{\xi} \nonumber\\
  &=&\ \langle n^{i\uparrow}\rangle_{\xi}\,\langle n^{i\downarrow}\rangle_{\xi}\ +
  \ \langle c^{i\uparrow}\cbar^{i\downarrow}\rangle_{\xi}\,\langle \cbar^{i\uparrow}c^{i\downarrow}
  \rangle_{\xi}\ .
\label{double3}
\eeqa
Note 
\beqa
\langle \cbar^{i\uparrow}c^{i\downarrow}\rangle_{\xi}\ &=&\ \half\bigl[\,\langle m^i_x\rangle_{\xi}\,
  +\,i\langle m^i_y\rangle_{\xi}\,\bigr] \nonumber\\
  \langle c^{i\uparrow}\cbar^{i\downarrow}\rangle_{\xi}\ &=&\ -\half\bigl[\,\langle m^i_x\rangle_{\xi}
  \,-\,i\langle m^i_y\rangle_{\xi}\,\bigr] \ ,
\label{double4}
\eeqa
while
\beqa
\langle n^{i\uparrow}\rangle_{\xi}\ &=&\ \half\bigl[\,\langle n^i\rangle_{\xi}\,+\,
  \langle m^i_z\rangle_{\xi}\, \bigr] \nonumber\\
\langle n^{i\downarrow}\rangle_{\xi}\ &=&\ \half\bigl[\,\langle n^i\rangle_{\xi}\,-\,
  \langle m^i_z\rangle_{\xi}\,\bigr]\ .
\label{double5}
\eeqa
Thus
\beq
\langle D^i\rangle_{\xi}\ =\ \fourth\bigl[\,\langle n^i\rangle^2_{\xi} \ -\ \langle
   \vecm^i\rangle_{\xi}\cdot\langle\vecm^i\rangle_{\xi}\,\bigr]\ .
\label{double6}
\eeq
The double occupancy assumes an explicitly spin--rotation--invariant form as required.

The expression in Eqn.~\ref{double6} should in principle be supplemented by zero-- and
finite--frequency RPA corrections to insure that Eqn.~\ref{double3} reproduces the $\tau\rightarrow
\tau'$ limit of the general two--body correlation function. (As noted below in the discussion of
response functions, RPA corrections appear at zero frequency because the charge interaction is
being treated within Hartree--Fock. There is no zero--frequency RPA in the spin channel, since
that portion of the interaction is being treated exactly.) RPA corrections in equal--time
expectation values $\langle n^{i\alpha}(\tau)n^{i\alpha'}(\tau)\rangle$ are actually problematic
in any perturbation theory for the following reason: The Pauli Principle implies an exact
identity
\beq
\langle n^{i\alpha}(\tau)n^{i\alpha}(\tau)\rangle\ =\ \langle n^{i\alpha}\rangle\ .
\label{double7}
\eeq
In perturbation theory, this identity can only be preserved exactly in approximations with a
crossing--symmetric two--body vertex. Such a vertex appears in order--by--order perturbation
theory (and in parquet theory), but not in a standard RPA, whether about the unperturbed vacuum
or about broken--symmetry states as in the present case. The RPA can be augmented to make it 
crossing--symmetric by incorporating the missing contributions in the crossed channel (see
Figure 3), but the result is no longer conserving in the Baym--Kadanoff sense.
The RPA corrections to the double occupancy are in any event expected to be small in the present
case. Since they are also problematic, we omit them from the calculation of $\langle D\rangle$
reported here.

We next consider the static spin and charge susceptibilities $\chi_s(Q)$ and $\chi_c(Q)$.
As mentioned above, one expects the appearance of RPA corrections in the static charge
response, since the charge interaction is being treated by ``slave'' Hartree--Fock, but
naively one might expect no corrections in the spin response. This is not correct: An average
must be performed over the symmetry--breaking field $\xii$, and in the presence of $\xii$
spin and charge excitations are mixed. For this reason the repulsive charge interaction makes its
presence felt weakly in the spin channel, as well as in the charge channel.

The static spin response function takes the form
\beq
\overline\chi^{ij}_{\mu\nu}\ \equiv\ \partial\,\langle\,\langle m^i_{\mu}\rangle_{\xi}\,\rangle\,
/\,\partial h^j_{\nu}\,\,\biggr|_{h=0}\ ,
\label{chi1}
\eeq
where the assumed local coupling to the external field $\vec h^j$ is
\beq
\Delta H\ =\ -\vec h^j\cdot\,\sum_{\alpha\alpha'}\,c^{\dag}_{j\alpha}\vec\sigma_{\alpha\alpha'}
  c^{\vphantom{\dag}}_{j\alpha'}\ .
\label{chi2}
\eeq
Temporarily neglecting RPA corrections, the field derivative yields
\beq
\overline\chi^{ij}_{\mu\nu}\ =\ -{1\over T}\,\bigl\langle\,\langle m^i_{\mu}\rangle_{\xi}\,
\bigr\rangle\,
  \bigl\langle\,\langle m^j_{\nu}\rangle_{\xi}\,\bigr\rangle 
  \ +\ {1\over T}\Bigl\langle\,
  \Bigl\langle \sum_n\,\cbar^{a\alpha}_n\sigma^{\mu}_{\alpha\alpha'}c^{i\alpha'}_n\,
  \sum_{n'}\,\cbar^{i\beta}_{n'}\sigma^{\nu}_{\beta\beta'}c^{j\beta'}_{n'}\Bigr\rangle_{\xi}\,\Bigr
  \rangle\ .
\label{chi3}
\eeq
The first term on the right--hand--side vanishes for zero applied field (since the system has no
spontaneous magnetization at finite temperature), and the second term (see Figure 4) reduces to
\beq
\overline\chi^{ij}_{\mu\nu}\ =\ {1\over T}\,\Bigl\langle\,\langle m^i_{\mu}\rangle_{\xi}\,
  \langle m^j_{\nu}\rangle_{\xi}\,\Bigr\rangle\ +\ \Bigl\langle\,
  -T\sum_n\sum_{\alpha\alpha'\beta\beta'}
  G_{\beta\alpha}(ji;\,i\omega_n)_{\xi}\,G_{\alpha'\beta'}(ij;\,i\omega_n)_{\xi}
  \,\sigma^{\mu}_{\alpha\alpha'}\,\sigma^{\nu}_{\beta'\beta}\,\Bigr\rangle\ ,
\label{chi4}
\eeq
i.e., formally an ``anomalous'' and a ``normal'' contribution. The anomalous contribution may be
evaluated using previous expressions for the site spin polarizations, while the
normal term may be written as the configurational average  of
\beq
\sum_{\alpha\alpha'\beta\beta'}\,M^{ij}_{\alpha\alpha',\,\beta\beta'}(i\Omega=0)_{\xi}
  \,\sigma^{\mu}_{\alpha\alpha'}\,\sigma^{\nu}_{\beta'\beta}\ ,
\label{chi5}
\eeq
where
\beq
M^{ij}_{\alpha_1\alpha_2,\,\beta_1\beta_2}(i\Omega)_{\xi}\ \equiv\ -\sum_{rs}\,\langle 
  j\beta_1|r\rangle\,\langle r|i\alpha_1\rangle\,\langle i\alpha_2|s\rangle\,
  \langle s|j\beta_2\rangle\,{ f(\varepsilon_r)-f(\varepsilon_s) \over \varepsilon_r
  +i\Omega-\varepsilon_s}\ .
\label{chi6}
\eeq
(Note that for $\varepsilon_r=\varepsilon_s$, the ratio involving the Fermi functions must be
replaced by $f'(\varepsilon_r)\delta_{\Omega 0}$. )

The static charge response function is
\beq
\overline\chi^{ij}\ \equiv\ \partial\,\langle\,\langle n^i\rangle_{\xi}\,\rangle\,
/\,\partial \mu^j\,\,\biggr|_{\mu=0}\ ,
\label{chi7}
\eeq
where in this case the assumed coupling to the external field is
\beq
\Delta H\ =\ -\mu^j\sum_{\alpha}\,c^{\dag}_{j\alpha}c^{\vphantom{\dag}}_{j\alpha}\ .
\label{chi8}
\eeq
In complete analogy with the spin response calculation the result before the inclusion of
RPA corrections is
\beqa
\overline\chi^{ij}\ &=&\ {1\over T}\,\Bigl[\,\Bigl\langle\, \langle n^i\rangle_{\xi}\,\langle n^j
  \rangle_{\xi}\,\Bigr\rangle\ -\ \langle n\rangle^2\,\Bigr] \nonumber\\
  &\phantom{=}&\qquad\qquad +\ \sum_{\alpha\alpha'\beta\beta'}\,\Bigl\langle M^{ij}_{\alpha\alpha',\,
  \beta\beta'}(i\Omega=0)_{\xi}\Bigr\rangle\,\delta_{\alpha\alpha'}\delta_{\beta'\beta}\ ,
\label{chi9}
\eeqa
with the normal ``bubble'' $M$ defined as in Eqn.~\ref{chi6}.

As emphasized above, RPA corrections necessarily appear when an interaction is treated 
perturbatively by Hartree--Fock. In the present case, where spin and charge excitations mix
before the configurational average is performed, RPA charge vertex corrections appear in both
the static charge and spin response functions. 

The calculation of RPA corrections for {\sl finite--frequency} response functions is trickier
than in conventional perturbation theory because the zero--frequency spin vertices have been
explicitly removed from the theory. This affects response functions at $i\Omega\neq 0$ by
partially removing vertices which carry zero frequency in a crossed particle--hole channel.
The analytic continuation to real energies is also complicated by this feature of the theory.
We defer discussion of spin and charge response at finite frequency and concentrate in the
present paper on RPA corrections to the static susceptibilities.

The repulsive RPA charge vertex produces a geometric series of corrections as in conventional
perturbation theory. Due to the lack of symmetry in individual configurations, the series can
only be summed in matrix form. Writing
\beq 
V^{ij}_{\alpha_1\alpha_2,\,\beta_1\beta_2}\ =\ \half U\delta_{ij}\delta_{\alpha_1\alpha_2}
  \delta_{\beta_2\beta_1}\ ,
\label{chi10}
\eeq
the complete RPA--corrected expressions for the spin and charge susceptibilities are given by
\beqa
\chi^{ij}_{\mu\nu}\ &=&\ {1\over T}\,\Bigl\langle\,\langle m^i_{\mu}\rangle_{\xi}\,\langle m^j_{\nu}
  \rangle_{\xi}\,\Bigr\rangle\ +\ \sum_{\alpha\alpha'\beta\beta'} \Bigl\langle\,\Bigl[\, M(1+VM)^{-1}\,
  \Bigr]^{ij}_{\alpha\alpha',\,\beta\beta'}(i\Omega=0)_{\xi}\,\Bigr\rangle\,\sigma^{\mu}
  _{\alpha\alpha'}\,\sigma^{\nu}_{\beta'\beta} \nonumber\\
  &\phantom{=}&
\label{chi11}
\eeqa
and
\beqa
\chi^{ij}\ &=&\ {1\over T}\,\Bigl[\,\Bigl\langle\,\langle n^i\rangle_{\xi}\,\langle n^j
  \rangle_{\xi}\,\Bigr\rangle\ -\ \langle n\rangle^2\,\Bigr] \nonumber\\
  &\phantom{=}&\qquad\qquad +\ \sum_{\alpha\alpha'\beta\beta'} \Bigl\langle\,\Bigl[\, M(1+VM)^{-1}\,
  \Bigr]^{ij}_{\alpha\alpha',\,\beta\beta'}(i\Omega=0)_{\xi}\,\Bigr\rangle\,\delta_{\alpha\alpha'}
  \delta_{\beta'\beta} \ .
\label{chi12}
\eeqa
Finally the momentum--resolved spin susceptibility takes the form
\beq
\chi_{\mu\nu}(Q)\ =\ {1\over\nsites}\sum_{ij}\,e^{iQ\cdot(R_i-R_j)}\,\chi^{ij}_{\mu\nu}\ ,
\label{chi13}
\eeq
and similarly for the charge susceptibility. (Measurements of the RPA--corrected susceptibilities
require the inversion of small matrices, but the effect on overall calculational efficiency is
negligible.)
\bigskip\bigskip
\section{Description of the Monte Carlo algorithm}

The effective action in Eqn. \ref{expect4} has been simulated using the Metropolis
algorithm. In comparison with a classical Heisenberg spin model, the present theory has several
complications: (i)~The theory has three degrees of freedom (orientation on the unit sphere and
amplitude) at each site, rather than two. (ii)~Local field updates induce global changes in
the action through the Fermi partition function $Z_{\rm f}(\xi)$. (iii)~A Hartree--Fock convergence
step is necessary after each spin field update to keep the ``slave'' charge fields 
$\langle n^i\rangle_{\xi}$ on the saddle point. (iv)~Single--site updating (i.e., Glauber dynamics)
is insufficient to achieve efficient equilibration for $\langle n\rangle\neq 1$. An analog of
Kawasaki dynamics, or spin swapping, is required, since the doping $1-\langle n\rangle$
plays a role similar to a conserved order parameter.

We consider each complication in turn. (i) The natural physical interpretation of field 
fluctuations in terms of
amplitude and orientation suggests an optimal representation for the functional integral over
field configurations. At half--filling it is expected that the only important 
fluctuations for $T\rightarrow
0$ correspond to magnons, i.e., slow modulations of the field orientation for fixed amplitude. A
Cartesian representation of the fields $(\xi^i_x,\,\xi^i_y,\,\xi^i_z)$ with independent
component updates leads to a small acceptance ratio in the low--temperature limit. Therefore
we write
\beq
\int\,d\xi^i_x\,d\xi^i_y\,d\xi^i_z\ =\ \int_0^{\infty}\,(r^i)^2\,dr^i\,\int_{-1}^1\,
d(\cos\theta^i)\,\int_0^{2\pi}\,d\phi^i\ ,
\label{measure1}
\eeq
choosing as independent variables $r^i$, $x^i\equiv\cos\theta^i$, and $\phi^i$. Variations in
the measure $(r^i)^2$ at each site are incorporated into the Metropolis updating factor, and
moves are restricted, in a way consistent with detailed balance, to lead to $r^i>0$. Proposed
changes to $r^i$ are restricted to a maximum scale $\Delta r$, while large--scale changes are
permitted in $x^i$ and $\phi^i$.

(ii) For the initial Monte Carlo calculations reported here we have simply diagonalized the
effective Hamiltonian matrix $H(\xi)$, which has row dimension $2\nsites$, for each
proposed $\vec\xi$ configuration. The time for a single site update thus grows
as the cube of the system size, and the time for a lattice sweep as the fourth power of the
system size. We believe that a hybrid molecular--dynamics/Monte--Carlo algorithm 
can be developed to significantly reduce the time needed for the simulation of large 
lattices \cite{sugar}.

(iii) The Hartree--Fock treatment of the static charge interaction is an integral component of
the present approach. More sophisticated treatments of fluctuations would also in principle
involve a self--consistent calculation in the presence of symmetry--breaking background fields.
For the parameter sets studied in the present paper, the Hartree--Fock iteration loop generally
converges in three to four steps. Since each step requires a matrix rediagonalization, overall
calculational time scales with the average number of Hartree--Fock iterations.

(iv) At half--filling the Hartree--Fock charge--field convergence becomes trivial,
and $\langle n^i\rangle$ remains equal to unity at all sites in all configurations. This just
reflects the particle--hole symmetry of the model for this special parameter set. Away from
half--filling the situation is quite different. Typical configurations contain a few sites with
small fields, while the majority of sites have fields with amplitudes close to the half--filled
value. Conventional single--spin updating tends to leave ``holes'' trapped at isolated locations,
since moving a hole requires passing through high--energy intermediate configurations.  In this
sense the number of hole $1-\langle n\rangle$ functions like a conserved order parameter,
preventing the equilibration of the system by Glauber dynamics. To combat this effect we have
introduced a second type of updating move, viz., a swap of the spin fields $\vec\xi$ at two sites
chosen at random. When the spin fields are swapped, the ``slave'' charge fields are dragged along,
and the system equilibrates to a uniform set of site occupancies $\langle\,\langle n^i\rangle_{\xi}\,
\rangle$, even though the hole distribution in individual configurations is highly non--uniform.
\bigskip
\bigskip

\section{Numerical results for a small Hubbard lattice}

The results for $4\times 4$ periodic lattices reported below were obtained on an IBM PowerPC
work station and a Sun--3000 server. For all runs reported here, a total of 100 Monte Carlo
warmup sweeps and 5000 sampling sweeps were carried out. Runs on $8\times 8$ and $10\times 10$ 
lattices are now in progress on a Sun--10000 work station array at the San Diego Supercomputer Center.

The total density of states for spin $\alpha$, $N_{\alpha}(\omega)$, is plotted in Figure
5 for $U/t=4$ and in Figure 6 for $U/t=8$, both at temperature $T/t=0.125$. The chemical 
potential $\mu/t$ is lowered from $U/2$ (half--filling) as holes are doped into
the insulating state. The mean site occupancy $\langle n\rangle$ and double occupancy 
$\langle D\rangle$ are summarized in Table 1. Note that the double occupancy 
at half--filling is reduced by factors of
order two and six from the value 0.25 for a non--interacting system. A prominent Mott--Hubbard
gap, or pseudogap, appears in the density of states for both parameter sets at half--filling
(Figures 5a and 6a). The remnants of the discrete quantum energy levels in the non--interacting 
system remain visible. Note again that
these systems are not magnetically ordered: The spin--up and spin--down densities of states
are identical within statistical sampling errors, and the spin susceptibility is isotropic, 
as discussed below. 

As the chemical potential is pulled below $\mu=U/2$, the site occupancy changes slowly
at first, particularly for the large charge gap which appears when $U/t=8$ (Figures 6b--c). 
This is just what one would expect for rigid--band doping in a semiconductor, and such behavior has
been observed in previous Lanczos \cite{dagotto1,dagotto2} and quantum Monte Carlo \cite{bulut}
studies of the Hubbard model on small lattices \cite{dagotto3}. Note
that the rigid--band picture begins to 
break down as soon as the site occupancy drops appreciably below unity (Figure 5b).
In the doped system with $U/t=4$ and $\mu/t=1.2$, 
spectral weight has been drawn from the lower edge of the upper Hubbard band into the gap.
Due to this transfer of weight, the gap feature rapidly collapses with doping, though remnants
remain for a wide range of fillings (Figures 5c--d). By the time $\mu/t$ reaches 0.4 for $U/t=4$, 
the shape of the non--interacting band
has begun to re--emerge, though levels are still significantly broadened by the fluctuating 
spin background.

This broadening is especially evident in the evolution of the k--resolved spectral density for
points near the Fermi surface. Recall that a special symmetry of the $4\times 4$ lattice
induces an equivalence between the generally distinct k--points $(\pi,\,0)$ and $(\pi/2,\,\pi/2)$.
This means all points on the half--filled Fermi surface
have the same $A_{k\alpha}(\omega)$ for the small systems discussed here. To contrast the 
behavior of points near and far from the Fermi surface, we plot $A_{k\alpha}(\omega)$ 
for $k=(\pi,\,0)$
and $k=(\pi,\,\pi)$ at $U/t=4$ in Figures 7 and 8. While the $(\pi,\,\pi)$ point at the 
top of the band (Figure 8) has a sharp quasiparticle--like
spectral density whose profile changes only slightly with doping, 
the $(\pi,\,0)$ point behaves quite differently. At half--filling $A_{k\alpha}$
for this point is symmetrically split between the upper and lower Hubbard band edges, just
as one would expect in an antiferromagnetic Hartree--Fock solution \cite{swz}. 
Away from half--filling, the spectral density
remains exceptionally broad, reflecting the continued presence of strong short--range antiferromagnetic
correlations between the fluctuating spins in the doped system. As noted above in the discussion
of $N(\omega)$, remnants of the half--filled gap remain for a wide range of fillings. The
spectral broadening decreases for fillings far from unity, consistent with the
restoration of the non--interacting band profile (and presumably the recovery of Fermi liquid behavior).

A broadening of the spectral density is also present in perturbative studies,
including self--consistent--field methods such as FLEX \cite{bw}; however, the details 
of the broadening mechanism are quite different in the present case. FLEX and
phenomenological spin--fluctuation models \cite{schmalian,monthoux} assume a small--amplitude scattering 
mechanism which may be
treated by perturbation theory about a non--interacting (i.e., Fermi liquid) background. In contrast
the present approach treats scattering by large--amplitude background distortions, with highly
non--Gaussian static fluctuations in both amplitude and orientation; these distortions evolve 
smoothly into the ordered antiferromagnetic background at half filling and zero temperature. 

The behavior of the spectral density at the $(\pi,\,0)$ point for $U/t=8$ is illustrated 
in Figure 9. As expected, the spectral weight is split symmetrically at half--filling, 
and the two peaks are separated by an energy of order $U$. In contrast with
the $U/t=4$ result, the weight is distributed broadly throughout the upper and lower Hubbard bands.
Signs of a more narrow coherent feature in the doped lower band are present for $\mu/t=1.8$
(Figure 9c).

The momentum dependence of the static charge and spin susceptibilities is plotted in Figures 10 and
11 for $U/t=4$ and 8 at temperature $T/t=0.125$. Note that the uniform charge susceptibility (or
compressibility) is exponentially small at half--filling, as expected for a system with a charge
gap. In addition the compressibility increases very rapidly with doping: Note particularly 
the increase for $U/t=8$ (Figure 11a). For general fillings $\chi_c(Q)$ is relatively
small (note the difference in scales for the charge and spin susceptibility) and structureless at this
temperature. This is expected in such a small system.
Possible manifestations of charge ordering within the fluctuating spin background may become
apparent, however, in larger systems.

In contrast the spin susceptibility is large and strongly peaked at the antiferromagnetic
wave vector $(\pi,\,\pi)$. The peak is expected to move off $(\pi,\,\pi)$ for doped systems 
on larger lattices. For the parameter sets illustrated here, the spin susceptibility is
dominated by the contributions of the ``anomalous'' term
\beq
{1\over\nsites}\sum_{ij}\,e^{iQ\cdot(R_i-R_j)}\,{ \Bigl\langle\,\langle m^i_{\mu}\rangle_{\xi}\,
  \langle m^j_{\nu}\rangle_{\xi}\,\Bigr\rangle \over T}\ .
\label{anom}
\eeq
\noindent
The $\mu\nu=zz$ component of $\chi_s$ is actually plotted in Figures 10 and 11.
For 5000 lattice sweeps the agreement between the $xx$, $yy$, and $zz$ components of the
$(\pi,\,\pi)$ spin susceptibility is at the level of the statistical error bars, i.e., a few percent. 
(The agreement can be arbitrarily improved by increasing the Monte Carlo sampling time.)
\bigskip\bigskip
\section{Extension to higher order Baym--Kadanoff approximations}

The present approach consists of simulation of the action for a set of local static fields
coupled to electron spin degrees of freedom. When the action is constrained to be
at its temperature--dependent global saddle point (if such a point indeed exists in the most
general case), these static fields are determined by inhomogeneous Hartree--Fock equations
and reduce to the anomalous fields of conventional diagrammatic perturbation theory. Just
as Hartree--Fock theory with an anomalous self--energy may be extended using the Baym--Kadanoff
prescription \cite{baym}, there is no formal barrier preventing such an extension in the present case.
The computational cost of carrying out self--consistent--field (SCF) perturbation theory
in a broken--symmetry background is quite large, however, since total momentum and spin are
no longer good quantum numbers. In this section we sketch out the steps necessary for a
generalized approach and note several of the technical pitfalls which must be studied and
overcome.

The form for the SCF action must be determined separately for each configuration of the
$\xii$ fields. In full analogy with Baym--Kadanoff theory for a uniform system, the action
takes the form
\beq
S^{\rm SCF}(\xi)\ =\ \beta\sum_i\,|\xi^i|^2\ -\ {\rm Tr}\,\log\,\bigl(-G^{\rm SCF}_{\Delta}
  \bigr)^{-1}\ +\ \beta\overline{\cal F}\ ,
\label{actionfourteen}
\eeq
where
\beq
\overline{\cal F}(\xi)\ =\ -{\rm Tr}\,\bigl(\Sigma G^{\rm SCF}\bigr)\ +\ \Phi
\label{freeenergy}
\eeq
and
\beq
{\delta\Phi\over \delta G^{\rm SCF}_{\sigma\sigma'}(xx')}\ =\ \Sigma_{\sigma'\sigma}(x'x)\ .
\label{phi}
\eeq
In these formal expressions $G^{\rm SCF}_{\Delta}$ is the discrete--time propagator
(retained to insure a proper high--frequency regularization of the ${\rm Tr}\,\log$ term);
\beq
x\ \equiv\ (\vec r^i,\,\tau)\ ,
\label{xx}
\eeq
and $\Phi$ is the SCF generating functional (e.g., a sum of ring and ladder diagrams in FLEX).

Since the SCF action may be interpreted as $\beta{\cal F}^{\rm SCF}$, with ${\cal F}^{\rm SCF}$
a well--behaved free energy, the statistical weighting factor $\exp(-S^{\rm SCF})$ 
is positive definite. This is in sharp contrast to the behavior in imaginary--time quantum
Monte Carlo simulations, where the weighting factor can be positive or negative with
nearly equal probability.

What is not so clear is the nature of approximations which insure the stability of
$S^{\rm SCF}$, particularly for exponentially rare field configurations. The interaction which
enters $S^{\rm  SCF}$  is not the original Hubbard interaction, since the zero--frequency
spin components have been removed for special treatment. This means that there are no
obvious zero--frequency instabilities left to appear in $S^{\rm SCF}$. Behavior at finite
frequencies must also be addressed, however, and we are not prepared to comment on the stability of
general Baym--Kadanoff approximations (and in particular FLEX) in the present paper.

As an example of the Baym--Kadanoff prescription in practice, note that when the static
charge interaction is treated within Hartree--Fock
\beq \Sigma_{\sigma\sigma'}(xx')\ =\ \half U\langle n^i\rangle_{\xi}\,\delta_{ii'}\delta_{\tau\tau'}
  \delta_{\sigma\sigma'}\ .
\label{sighf}
\eeq
The appropriate generating functional is
\beq
\Phi\ =\ \fourth U\sum_i\,\langle n^i\rangle_{\xi}^2\ .
\label{phii}
\eeq
Furthermore
\beq
{\rm Tr}\,(\Sigma G^{\rm SCF})\ =\ 2\Phi
\label{trace}
\eeq
and so
\beqa
\overline{\cal F}\ &=&\ -\Phi \nonumber\\
  &=&\ -\fourth U\sum_i\,\langle n^i\rangle_{\xi}^2\ .
\label{calf}
\eeqa
Note that this form for the Hartree--Fock ``constant'' in the action is exactly that which has
appeared previously in Eqn.~\ref{actiontwelve}.

\section{Some speculations and conclusions}

We believe the present Hubbard model analysis, based on Monte Carlo simulation of a classically
disordered local spin field, incorporates several desirable features which have eluded previous
perturbative studies. Most important the present theory has classical critical behavior at
half--filling and describes the formation of a Mott--Hubbard gap (or pseudogap) in a system
with a finite spin correlation length. In addition the spectral densities for the doped systems
have the features suggested by exact methods.

The present
static field analysis can be extended to significantly larger systems. Since the computations
involve Monte Carlo field sampling and have reasonably small memory requirements, it would be natural
to carry out studies on multiple processors with different random seeds, then to merge the
equilibrated results from numerous short runs.

It is almost unavoidable to speculate on the implications for charge density wave or stripe
formation \cite{kivelson,salkola} and d--wave singlet pairing in the cuprate superconductors. Since the present
approach reduces to standard inhomogeneous Hartree--Fock for $T\rightarrow 0$, Hartree--Fock results
previously obtained for stripe formation should re--emerge \cite{zaanen}. By allowing sampling of 
configurations removed from the zero--temperature saddle point in a well--defined way,
the present approach provides a framework to describe the destruction of striped order by
thermal fluctuations.

To study potential d--wave superconductivity it seems essential to incorporate the exchange of
low--frequency spin fluctuations about the disordered background configurations. The
low--energy fluctuations which become magnons in the directions transverse to
an ordered spin state (i.e., the Goldstone modes of the order parameter field) must evolve into
a broadened spectrum of fluctuations in the important disordered configurations. If low--frequency 
spin fluctuations are important,
the pair order parameter should exhibit retardation effects, in contrast
with the spin order parameter. In addition the disordered background has important implications for
the onset of pairing. Presumably the pair field appears initially with a highly 
non--uniform distribution in a small subset of $\vec\xi$ configurations. Such behavior would imply
superconducting precursor effects in the density of states well above the transition
temperature for global phase coherence. The effects would be more akin to the behavior of a
system of superconducting grains with variable grain size than to conventional Kosterlitz--Thouless 
phase fluctuations in a clean two--dimensional superconductor.

In any case the scenario described above is at this stage purely speculative. The next steps in
the development of the present approach consist of (i)~studies of larger systems; (ii)~analysis of
the spin and charge response at finite frequencies, with particular care given to the
treatment of potential instabilities in rare field configurations (see Section 6); and
(iii)~analysis of pairing eigenvalues and eigenvectors from exchange of fluctuations in
sets of non--uniform configurations.
\bigskip\bigskip\bigskip

We gratefully acknowledge conversations with S. Haas, W. Hanke, S.A. Kivelson, H. Monien, S. Moukouri, 
R.T.  Scalettar, R.L. Sugar and S.--C. Zhang during
the course of this work. NEB acknowledges support from the National Science 
Foundation under grant no. DMR98--11381, and DJS acknowledges support from the 
U.S. Department of Energy under grant no. DE--FG03--85ER45197.
\vfill\eject

\vskip 3.0in
\noindent
{\footnotesize Table 1. Mean occupancy $\langle n\rangle$ and 
double occupancy $\langle D\rangle$ for
parameter sets at $T/t=0.125$. Statistical errors are in the last digit.}
\bigskip\bigskip
\tabskip=1emplus2em minus.5em
$$\vbox{\halign to \hsize{\hfil#\hfil&\hfil#\hfil&\hfil#\hfil&\hfil#\hfil&\hfil#\hfil&\hfil#\hfil\cr

$\phantom{----}$ & $U/t$  & $\mu/t$ & $\langle n\rangle$ & $\langle D\rangle$ & $\phantom{----}$\cr
& \cr
& 4.0  & 2.0  & 1.00 & 0.118& \cr
& 4.0  & 1.2  & 0.93 & 0.105& \cr
& 4.0  & 1.0  & 0.88 & 0.095& \cr
& 4.0  & 0.8  & 0.83 & 0.086& \cr
& 4.0  & 0.6  & 0.77 & 0.078& \cr
& 4.0  & 0.4  & 0.72 & 0.071& \cr
& 8.0  & 4.0  & 1.00 & 0.037& \cr
& 8.0  & 2.4  & 0.99 & 0.037& \cr
& 8.0  & 1.8  & 0.95 & 0.033& \cr }}$$
\vfill\eject
\begin{figure}[htb]
\caption[]{Diagrammatic representation of the Hubbard interaction vertex.}
\vskip 1.5 in
\epsfxsize=4.0in
\hbox to\hsize{\hfil\epsfbox{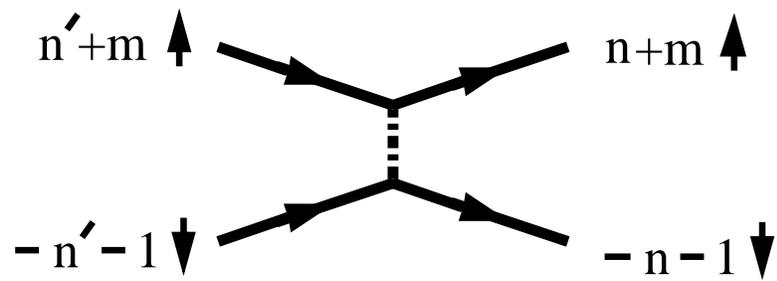}\hfil}
\end{figure}
\vfill\eject
\begin{figure}[htb]
\caption[]{Rewriting of the Hubbard interaction vertex in the two crossed
particle--hole channels. (a)~Transverse spin channel. (b)~Longitudinal spin and charge
channel.}
\vskip 1.5 in
\epsfxsize=4.0in
\hbox to\hsize{\hfil\epsfbox{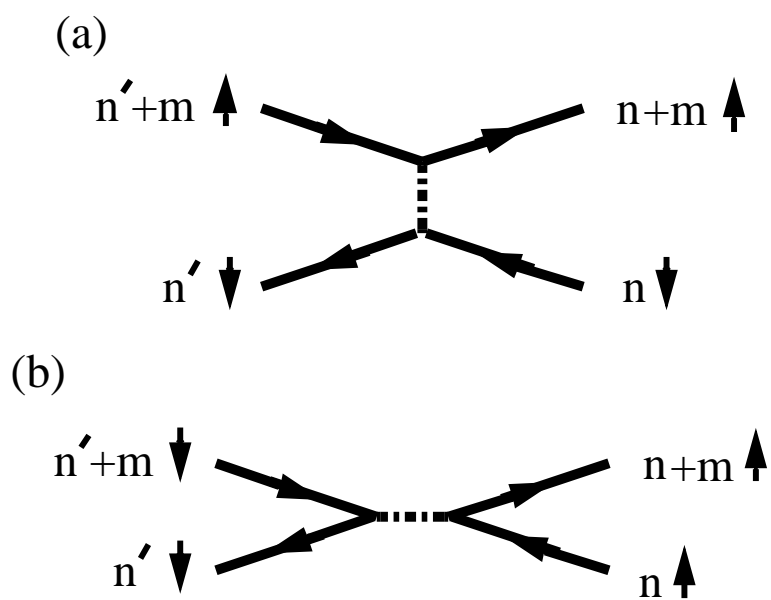}\hfil}
\end{figure}
\vfill\eject
\begin{figure}[htb]
\caption[]{Example of reducible and irreducible particle--hole vertex diagrams 
related by crossing symmetry. The third--order reducible vertex appears in an RPA
summation. It must be augmented
by the irreducible particle--hole vertex shown in order to preserve the Pauli Principle
identity in Eqn.~\ref{double7}.}
\vskip 1.5 in
\epsfxsize=4.0in
\hbox to\hsize{\hfil\epsfbox{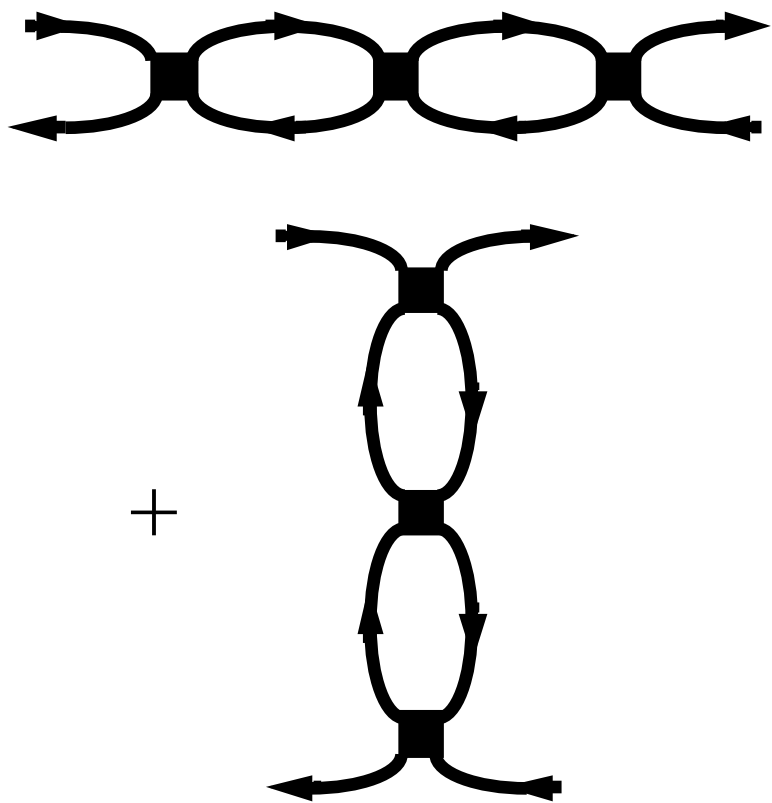}\hfil}
\end{figure}
\vfill\eject
\begin{figure}[htb]
\caption[]{ Contributions to the static spin response function. (a)~``Anomalous''
diagram. (b)~``Normal'' diagram.}
\vskip 1.5 in
\epsfxsize=4.0in
\hbox to\hsize{\hfil\epsfbox{Fig4.epsf}\hfil}
\end{figure}
\vfill\eject
\begin{figure}[htb]
\caption[]{
Total density of states for spin $\alpha$, $N_{\alpha}(\omega)$.
The interaction strength is $U/t=4$, and the temperature is fixed at $T/t=0.125$. Results
are shown for four different chemical potentials. See Table 1 for the mean site occupancy
and double occupancy values. Note that in the absence of statistical sampling error
$N_{\uparrow}(\omega)=N_{\downarrow}(\omega)$. Therefore the measured difference in
these two quantities provides a point--by--point error estimate. The vertical dashed line
at $\omega=0$ is drawn to emphasize the position of the chemical potential. (a)~$\mu/t=2.0$.
(b)~$\mu/t=1.2$. (c)~$\mu/t=0.8$. (d)~$\mu/t=0.4$.}
\vskip 1.5 in
\epsfxsize=6.0in
\hbox to\hsize{\hfil\epsfbox{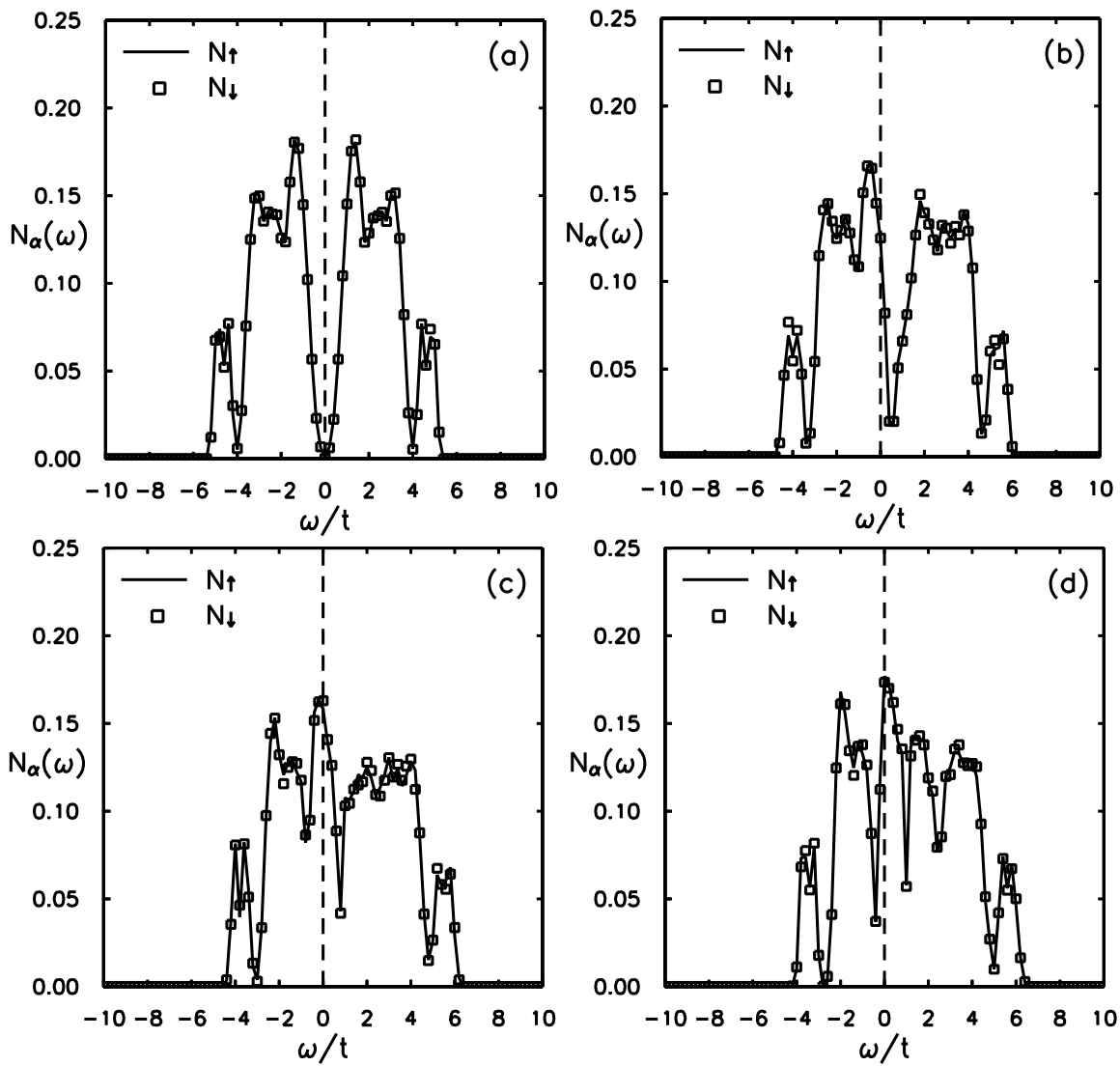}\hfil}
\end{figure}
\vfill\eject
\begin{figure}[htb]
\caption[]{Total density of states for spin $\alpha$, $N_{\alpha}(\omega)$.
The interaction strength is $U/t=8$, and the temperature is fixed at $T/t=0.125$, as in
Figure 5. See Table 1 for the mean site occupancy and double occupancy values. (a)~$\mu/t=4.0$.
(b)~$\mu/t=2.4$. (c)~$\mu/t=1.8$.}
\vskip 1.5 in
\epsfxsize=6.0in
\hbox to\hsize{\hfil\epsfbox{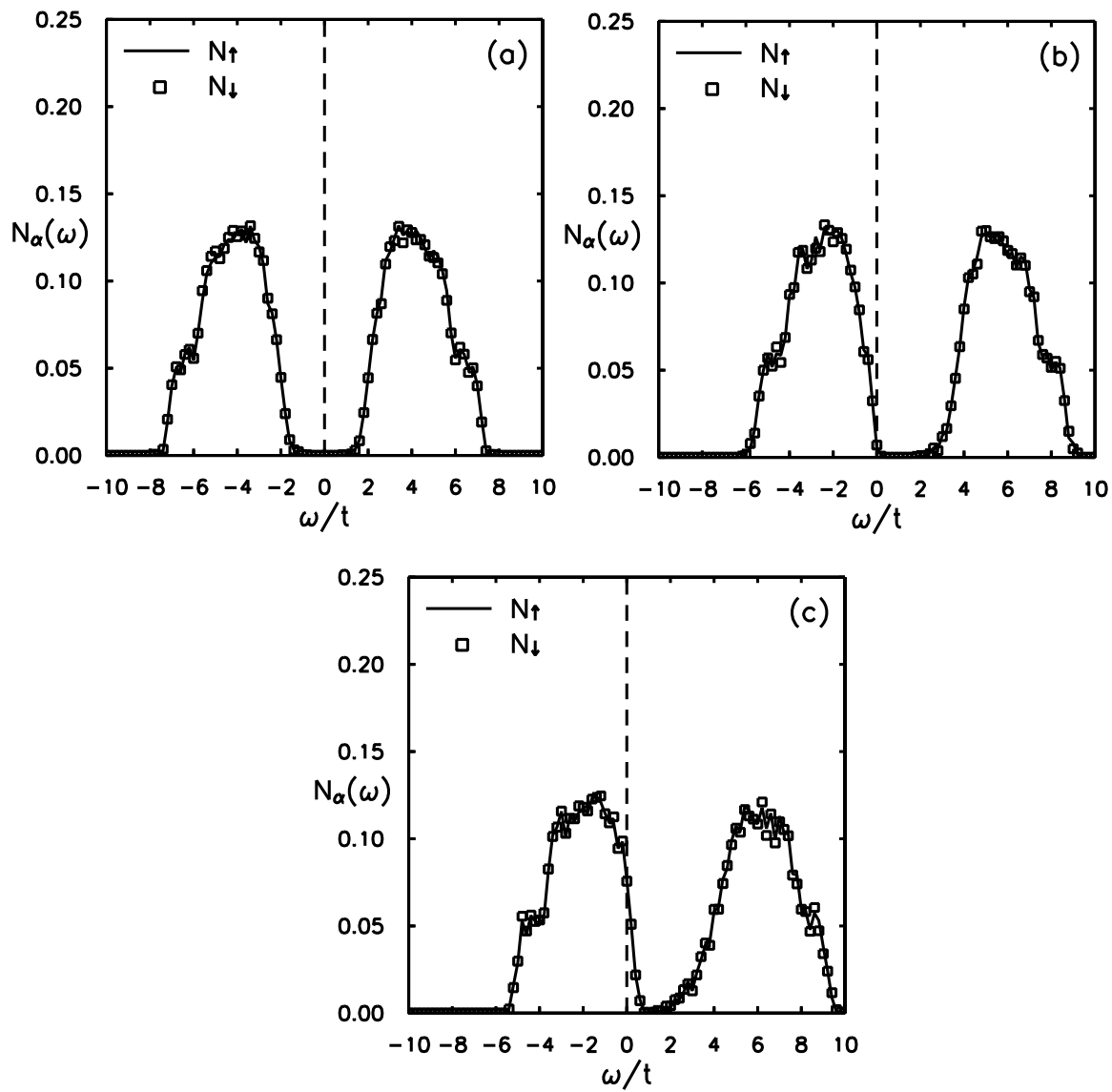}\hfil}
\end{figure}
\vfill\eject
\begin{figure}[htb]
\caption[]{Momentum--resolved spectral density $A_{k\alpha}(\omega)$.
The model parameters are $U/t=4$ and $T/t=0.125$, as in Figure 5. The momentum point chosen
is $k=(\pi,\,0)$. (a)~$\mu/t=2.0$. (b)~$\mu/t=0.8$. (c)~$\mu/t=0.4$.}
\vskip 1.5 in
\epsfxsize=6.0in
\hbox to\hsize{\hfil\epsfbox{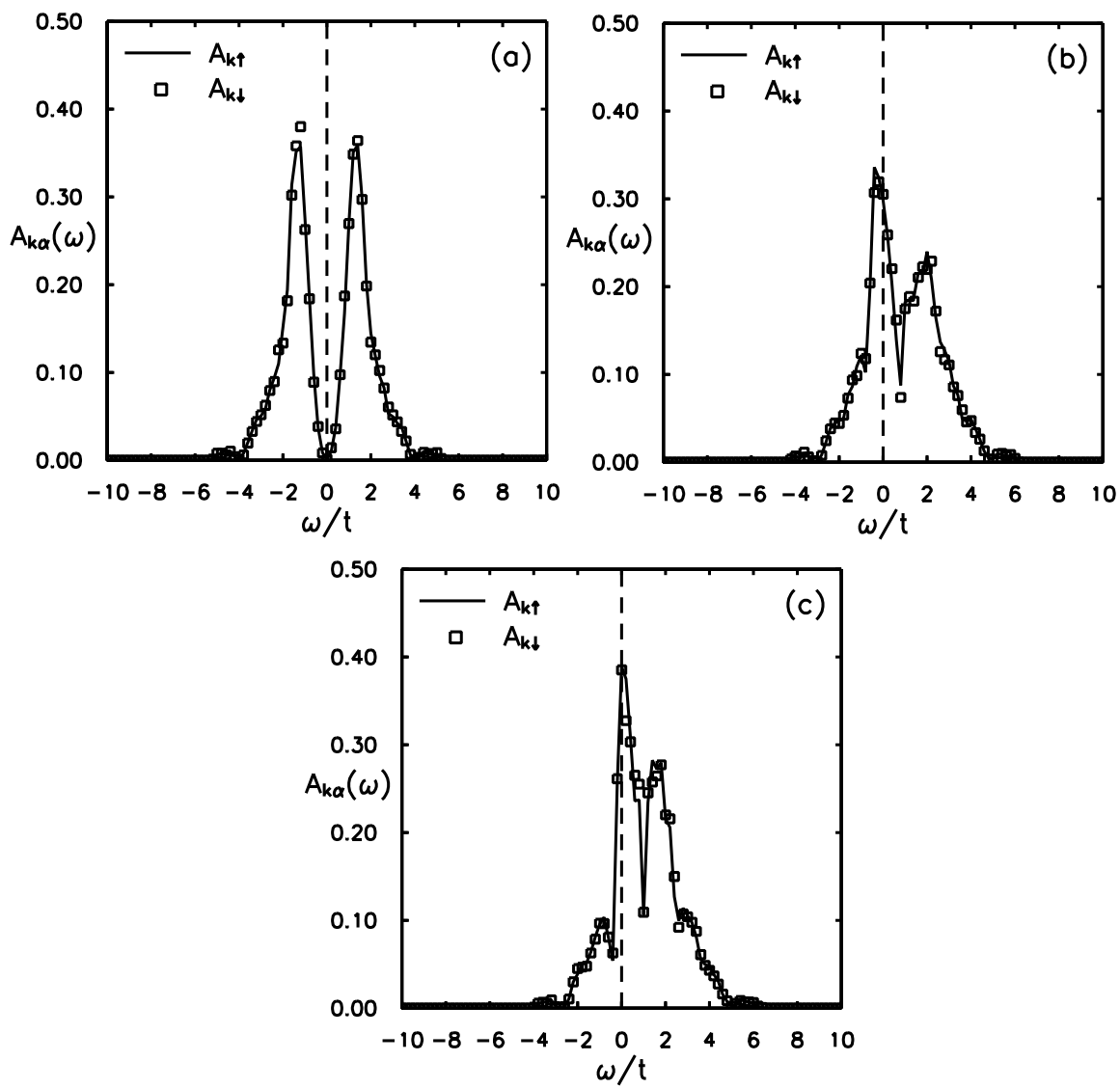}\hfil}
\end{figure}
\vfill\eject
\begin{figure}[htb]
\caption[]{Momentum--resolved spectral density $A_{k\alpha}(\omega)$.
The model parameters are as in Figure 5 and 7. The momentum point chosen
is $k=(\pi,\,\pi)$. (a)~$\mu/t=2.0$. (b)~$\mu/t=0.4$.}
\vskip 1.5 in
\epsfxsize=6.0in
\hbox to\hsize{\hfil\epsfbox{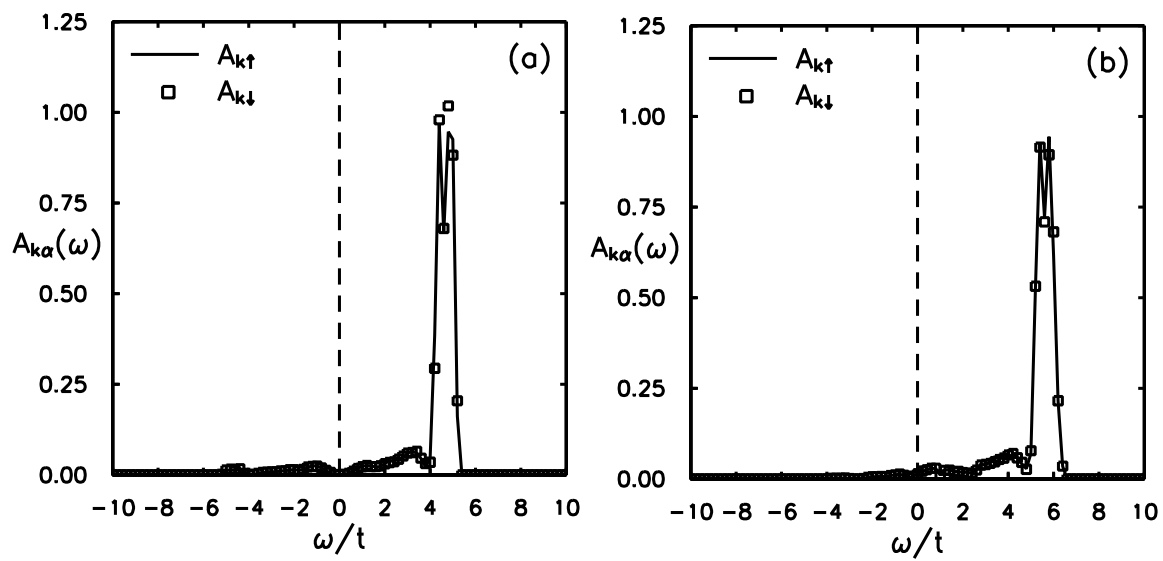}\hfil}
\end{figure}
\vfill\eject
\begin{figure}[htb]
\caption[]{Momentum--resolved spectral density $A_{k\alpha}(\omega)$.
The model parameters are $U/t=8$ and $T/t=0.125$, as in Figure 6. The momentum point chosen
is $k=(\pi,\,0)$. (a)~$\mu/t=4.0$. (b)~$\mu/t=2.4$. (c)~$\mu/t=1.8$.}
\vskip 1.5 in
\epsfxsize=6.0in
\hbox to\hsize{\hfil\epsfbox{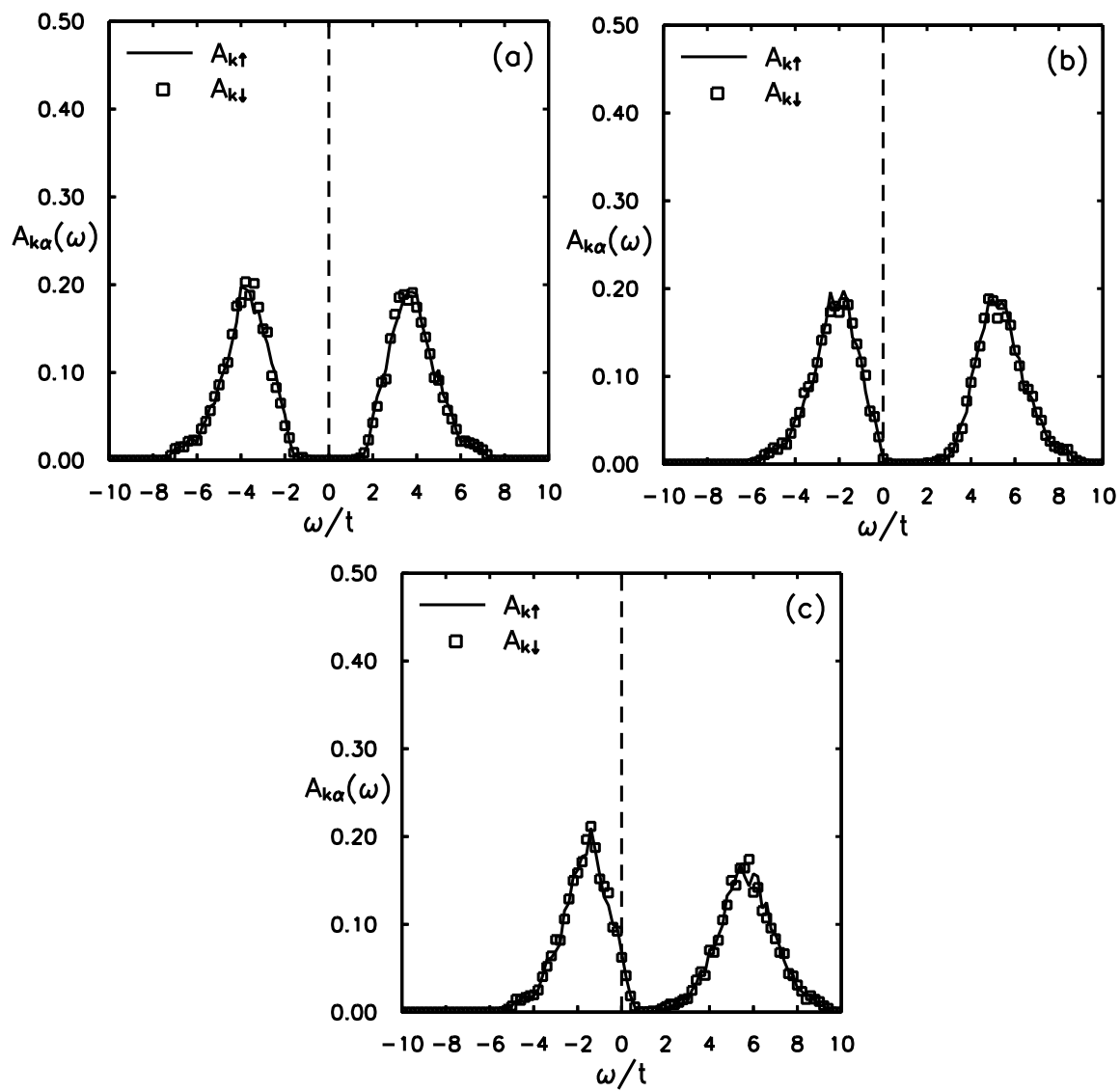}\hfil}
\end{figure}
\vfill\eject
\begin{figure}[htb]
\caption[]{Static charge susceptibility $\chi_c(Q)$ and spin susceptibility $\chi_s(Q)$ for
$U/t=4$ and $T/t=0.125$. The
susceptibilities are plotted along the triangular Brillouin zone contour from $\Gamma\rightarrow X
\rightarrow M\rightarrow\Gamma$, i.e., from $(0,\,0)\rightarrow(\pi,\,0)\rightarrow(\pi,\,\pi)
\rightarrow(0,\,0)$. Results are shown for the chemical potentials and site occupancies
employed in Figure 5. Unless shown, statistical error bars are smaller than the plotting symbols.
(a)~$\chi_c(Q)$. (b)~$\chi_s(Q)$.}
\vskip 1.5 in
\epsfxsize=6.0in
\hbox to\hsize{\hfil\epsfbox{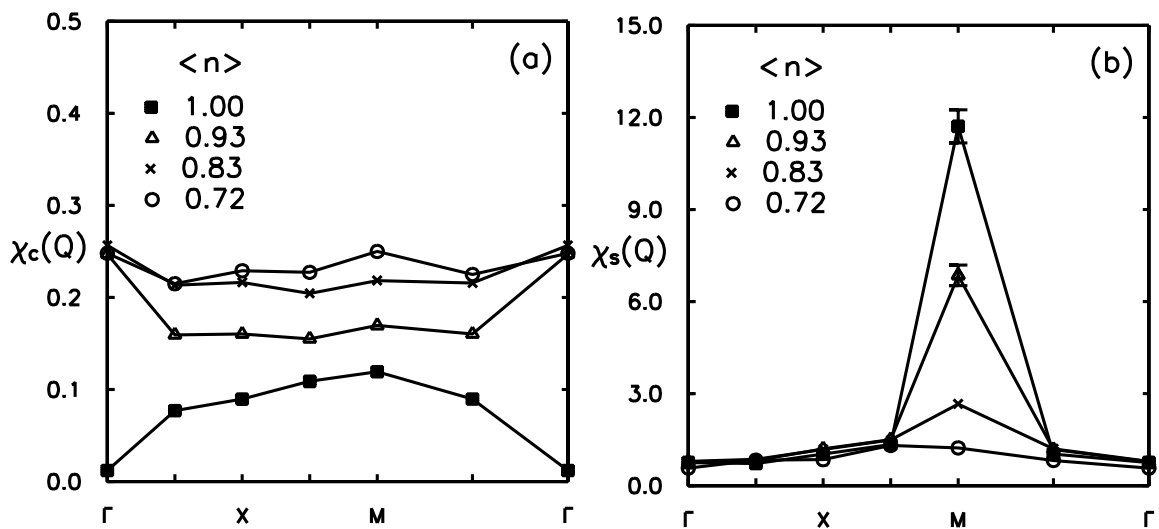}\hfil}
\end{figure}
\vfill\eject
\begin{figure}[htb]
\caption[]{Static charge susceptibility $\chi_c(Q)$ and spin susceptibility $\chi_s(Q)$ for 
$U/t=8$ and $T/t=0.125$. The
susceptibilities are plotted along the triangular contour used in Figure 10. Results
are shown for the chemical potentials and site occupancies employed in Figure 6.
(a)~$\chi_c(Q)$. (b)~$\chi_s(Q)$.}
\vskip 1.5 in
\epsfxsize=6.0in
\hbox to\hsize{\hfil\epsfbox{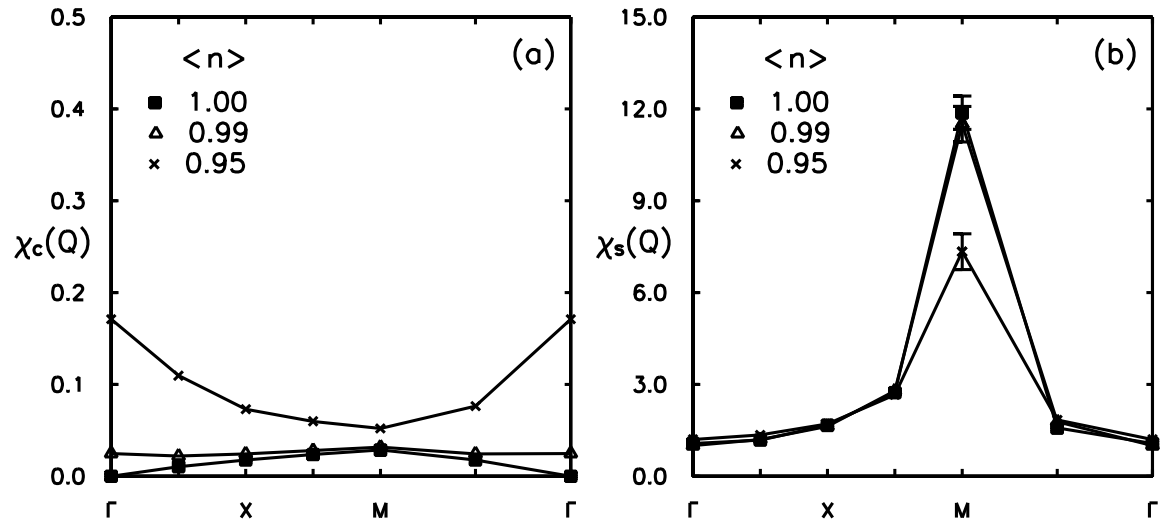}\hfil}
\end{figure}
\vfill\eject

\begin{thebibliography}{99}
\bibitem{correction} The $n'=n$ correction is present because equivalent components
in the order parameter field appear in crossed scattering channels. In contrast no such correction
appears when one decouples a general instantaneous pairing interaction or a Holstein phonon
interaction. While the $n'=n$ term must be retained to properly regain the high temperature
limit, it seems clear that it has no special physical implications at low temperatures. Further
the correction may at least formally be treated exactly by introducing a ``strong--coupling''
(but still $m=0$) generalization of the $\xii$ fields.
\bibitem{rough} It is possible to provide a heuristic derivation of a one--electron
self--energy with the general form in Eqn.~\ref{perturb} starting from the full theory for $S_{\rm eff}$
in Section 2. The steps are as follows: (i)~Write down the expression for the one--electron propagator
dictated by Eqn.~\ref{expect}. (ii)~Expand the expression for $G_{i\alpha,\,j\beta}(i\omega_n)_{\xi}$
within the $\xii$ integral as a power series in $\xii\cdot\vec\sigma$, noting that the odd--order
terms integrate to zero. (iii)~Resum the series within the integral as
$[G_0^{-1}(i\omega_n)-\Sigma(i\omega_n)_{\xi}]^{-1}$, where $\Sigma(i\omega_n)_{\xi}$ takes the form of
a self--energy second--order in $\xii$. (iv)~Approximate the resulting field integral 
as $[G_0^{-1}(i\omega_n)-\langle\Sigma(i\omega_n)_{\xi}\rangle]^{-1}$, i.e., average the
second--order self--energy, rather than the propagator. The field--averaged self--energy
assumes exactly the form of Eqn.~\ref{perturb}, with $V$ proportional to the full static spin
susceptibility of the system. The electron propagator which appears in the self--energy is bare 
for the reason discussed in the text. Step (iv) is of course not justifiable,
since the action is strongly non--Gaussian, but it provides the phenomenological link between
some previous theories and the current work.
\end{thebibliography}
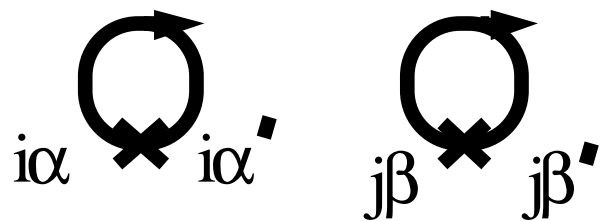
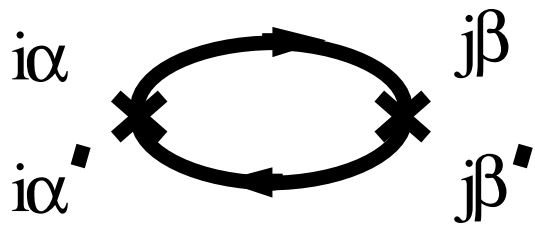
\end{document}